\documentclass[format=acmlarge, screen]{acmart}
\AtBeginDocument{%
  }

\setcopyright{acmlicensed}
\acmJournal{IMWUT}
\copyrightyear{2025}
\acmYear{2025}
\acmDOI{XXXXXXX.XXXXXXX}

\acmISBN{978-1-4503-XXXX-X/2018/06}
\usepackage{xspace}
\usepackage[utf8]{inputenc}
\usepackage{enumitem}
\usepackage{xcolor}
\usepackage{mdframed}
\usepackage{subcaption}
\newcommand{\highlight}[1]{\vspace{0.0cm}\begin{mdframed}[backgroundcolor=gray!15]#1\end{mdframed}}
\mdfsetup{skipabove=1.5em,skipbelow=1.5em}
\usepackage{hyperref}
\usepackage{multirow}
\usepackage{multicol}
\usepackage{color}
\usepackage{colortbl}
\definecolor{Gray}{gray}{0.9}
\newcolumntype{g}{>{\columncolor{Gray}}c}
\usepackage{graphicx}
\usepackage{makecell}
\usepackage{arydshln}

\begin{document}
\title{Computing with Smart Rings: A Systematic Literature Review}



\author{Zeyu Wang}
\email{wang-zy23@mails.tsinghua.edu.cn}
\orcid{0009-0007-5048-1665}
\affiliation{%
  \institution{Key Laboratory of Pervasive Computing, Ministry of Education, Department of Computer Science and Technology, Tsinghua University}
  \country{China}
}

\author{Ruotong Yu}
\email{yurt24@mails.tsinghua.edu.cn}
\orcid{0009-0006-3691-3110}
\author{Xiangyang Wang}
\orcid{0009-0001-5762-3431}
\email{xiangyang24@mails.tsinghua.edu.cn}
\affiliation{%
  \institution{Tsinghua University}
  \country{China}
}

\author{Jiexin Ding}
\email{jxding@uw.edu}
\affiliation{%
    \department{Electrical and Computer Engineering}
   \institution{University of Washington}
   \country{The United State}
}

\author{Jiankai Tang}
\orcid{0009-0009-5388-4552}
\author{Jun Fang}
\email{fangy23@mails.tsinghua.edu.cn}
\orcid{0009-0001-2614-8674}
\author{Zhe He}
\orcid{0000-0001-5874-1096}
\email{hz23@mails.tsinghua.edu.cn}
\author{Zhuojun Li}
\orcid{0000-0003-4374-9452}
\email{lizj23@mails.tsinghua.edu.cn}
\affiliation{%
  \institution{Tsinghua University}
  \country{China}
}

\author{Tobias R\"{o}ddiger}
\email{tobias.roeddiger@kit.edu}
\orcid{0000-0002-4718-9280}
\affiliation{%
  \institution{Karlsruhe Institute of Technology}
  \country{Germany}
}

\author{Weiye Xu}
\orcid{0000-0001-5031-8154}
\email{xuwy24@mails.tsinghua.edu.cn}
\author{Xiyuxing Zhang}
\email{zxyx22@mails.tsinghua.edu.cn}
\orcid{0009-0002-9337-2278}
\author{Huan-ang Gao}
\orcid{0009-0004-6727-5778}
\email{gha24@mails.tsinghua.edu.cn}
\author{Nan Gao}
\orcid{0000-0002-9694-2689}
\email{nangao@tsinghua.edu.cn}
\author{Chun Yu}
\orcid{0000-0003-2591-7993}
\email{chunyu@tsinghua.edu.cn}
\affiliation{%
  \institution{Tsinghua University}
  \country{China}
}

\author{Yuanchun Shi}
\orcid{0000-0003-2273-6927}
\affiliation{%
  \institution{Key Laboratory of Pervasive Computing, Ministry of Education, Department of Computer Science and Technology, Tsinghua University}
  \country{China}
}
\affiliation{%
  \institution{Intelligent Computing and Application Laboratory of Qinghai Province, Qinghai University}
  \country{China}
}
\email{shiyc@tsinghua.edu.cn}

\author{Yuntao Wang}
\authornote{Corresponding authors.}
\email{yuntaowang@tsinghua.edu.cn}
\orcid{0000-0002-4249-8893}
\affiliation{%
  \institution{Key Laboratory of Pervasive Computing, Ministry of Education, Department of Computer Science and Technology, Tsinghua University}
  \country{China}
}

\renewcommand{\shortauthors}{Zeyu Wang et al.}

\begin{abstract}
A smart ring is a wearable electronic device in the form of a ring that incorporates diverse sensors and computing technologies to perform a variety of functions. Designed for use with fingers, smart rings are capable of sensing more subtle and abundant hand movements, thus making them a good platform for interaction. Meanwhile, fingers are abundant with blood vessels and nerve endings and accustomed to wearing rings, providing an ideal site for continuous health monitoring through smart rings, which combine comfort with the ability to capture vital biometric data, making them suitable for all-day wear. We collected in total of 206 smart ring-related publications and conducted a systematic literature review. We provide a taxonomy regarding the sensing and feedback modalities, applications, and phenomena. We review and categorize these literatures into four main areas: (1) interaction - input, (2) interaction - output, (3) passive sensing - in body feature, (4) passive sensing - out body activity. This comprehensive review  highlights the current advancements within the field of smart ring and identifies potential areas for future research.
\end{abstract}



\maketitle

\section{Introduction}
Over the past years, wearable computing devices have gained increasing popularity such as smartwatches~\cite{king2018survey}, smart earables~\cite{roddiger2022sensing}, smart glasses~\cite{wang2024g} and smart rings~\cite{rissanen2013subtle}. 
Among these, smart rings are lightweight, socially
acceptable across cultures, and seamlessly integrate into daily life with minimal obtrusiveness~\cite{bilius2023could, rissanen2013subtle}.
However, despite their potential, there remains a gap in comprehensive understanding and systematic documentation of the capabilities and advancements in smart ring technology. This raises an important question: \textit{What are smart rings capable of, and how can they enable interaction, well-being, and overall lifestyle management?}

Depending on the wearing position, smart rings can capitalize on the unique properties of the human finger. Fingers are the termini of our limbs, equipped with rich sensory innervations and capable of intricate movements. 
Smart rings are positioned in a way that allows them to sense a wide range of motion-related phenomena, making them exceptionally good at detecting gestures~\cite{vatavu2021gesturing} and fine motor activities~\cite{rissanen2013subtle}. 
The flexibility of fingers enable a high degree of freedom (DOF) spatial interactions, which can be harnessed for complex command inputs in various applications, substituting numerous input devices, including mouse~\cite{chen2014mobiring, waghmare2023zring, shen2024mousering}, keyboard~\cite{liang2023drg, gu2020qwertyring, li2023ringvkb, nirjon2015typingring}, or controller~\cite{yau2020subtle, yeo2019wrist}. 
Moreover, due to the fingers' constant blood flow and exposure, smart rings can effectively monitor physiological signals that are typically checked in clinical settings, such as pulse rate~\cite{boukhayma2021ring, haddad2021ear, dong2021cloud, mahmud2018sensoring} or oxygen saturation~\cite{magno2019self}, thus enabling health tracking applications~\cite{rajput2023assessment, usman2019analyzing, halkola2019towards}.
Additionally, the strategic placement on the finger enables continuous health monitoring~\cite{wongtaweesup2023using, joseph2022integrated} without the need for more invasive or cumbersome equipment. 
This dual capability of interaction and passive sensing underscores the potential of smart rings as both input/output devices and health monitoring systems.

Prior to this work, several surveys related to smart ring technologies have been published. 
In 2013, \citet{rissanen2013subtle} conducted an early survey of ring-shaped user interfaces based on 16 research articles to illustrate the ``ringterface'' concept. 
In 2015, \citet{shilkrot2015digital} conducted a comprehensive survey on finger augmentation devices (FADs), covering sensing and feedback modalities on smart rings. Their work also included non-ring form factors, such as fingernail addendums and finger sleeves. 
However, given the significant advancements and increasing commercialization of smart rings over the past ten years, there is a pressing need for an updated survey that reflects the current state and emerging trends in smart ring technology. 
In the most recent work from 2021, \citet{vatavu2021gesturing} conducted a systematic literature review on ring-based gesture input and contributed a searchable gesture-to-function dictionary. 
However, there are numerous applications that can be enabled by a ring form-factor that go beyond just gesture recognition.  
Therefore, in comparison to previous works, our survey consolidates the full spectrum of capabilities enabled by smart rings, encompassing a broader range of functionalities to provide a comprehensive overview of the field's current state and future directions.

In this paper, we follow the FADs form factors proposed by~\citet{shilkrot2015digital} and solely focus on the ``rings'' category. 
We explore the research field of smart rings by conducting a systematic literature review of 206 smart ring-related publications. 
We present a taxonomy that categorizes the sensing and feedback modalities, applications, and the phenomena these devices can detect and interact with. Our review organizes the existing literature into four main areas: (1) interaction - input, where we explore how smart rings serve as command interfaces; (2) interaction - output, examining the feedback mechanisms rings can provide; (3) passive sensing - in-body features, detailing how physiological and biometric data are captured for health monitoring; and (4) passive sensing - out-body activity, discussing how rings monitor external physical activities and environmental cues. This comprehensive review highlights the advancements within the field and also identifies potential areas for future research.

\section{Review Methodology} \label{sec:methodology}
In our work we draw inspiration from processes employed in related \textit{Systematic Literature Reviews (SLR)}~\cite{ahmadilivani2024systematic, lavallee2013performing, roddiger2022sensing}. We (i) specify the research questions (RQs); (ii) specify a definition for smart rings; (iii) specify the search database and keywords to retrieve relevant research papers; (iv) perform backward chaining to expand the paper set; (v) analyze and extract content based on the RQs; and (vi) analyze the extracted contents to form a taxonomy for smart rings.
Based on the guidelines mentioned above, our work focuses on addressing the following research questions:
\begin{itemize}
    \item \textbf{RQ1:} What is the distribution of the smart ring literature?
    \item \textbf{RQ2:} What sensing methods have been employed in smart rings?
    \item \textbf{RQ3:} What smart ring applications are enabled by the sensing methods?
    \item \textbf{RQ4:} What are the challenges and future opportunities of smart rings?
\end{itemize}

To establish a working definition for \textit{smart rings}, we examined definitions proposed in related work. 
Gheran et al.~\cite{gheran2018gestures} and Vatavu et al.~\cite{vatavu2021gesturing} define ring gestures as \textit{``any action performed with or on a smart ring or any movement of the wearing finger and/or hand that causes a detectable change in the ring’s position and/or orientation in a system of reference centered on the user’s finger or body''}. 
In contrast to this interaction-centered definition, our survey seeks to cover the full breadth of smart ring applications including physiological and health sensing, authentication, and haptic feedback.
Vatavu et al.~\cite{vatavu2021gesturing} described wearing form-factors as "ring", "ring-like" and "ring-ready".  
In our work, we rely on the ``ring'' form-factor as described by Shilkrot et al.~\cite{shilkrot2015digital}. Therefore, we broadly define smart rings as follows:

\highlight{\textbf{Smart Rings} are small, circular electronic devices that are designed to be worn on a phalanx of a finger.}

Following the search strategy of related survey papers~\cite{brudy2019cross, ahmadilivani2024systematic, roddiger2022sensing, vatavu2021gesturing}, we perfomed a keyword-based search on the ACM Digital Library and IEEE Explore, which are believed to contain the majority of wearable and HCI publications. In total, we this results in 593 relevant publications with 199 from the ACM and 394 from the IEEE repositories.
The keywords we used for each database were as follows:

\highlight{

    \textbf{ACM Digital Library:}
    \begin{itemize}
        \item query target: Title, Keywords, Abstract
        \item keywords: finger-worn, finger-based, finger-mounted, finger-attached, ring AND [allfield] wearable
        \item applied filters: Research Article, Short Paper
    \end{itemize}

\vspace{0.4cm}

\noindent\textbf{IEEE Explore:}
    \begin{itemize}
        \item query target: Document Title, Index Terms, Abstract
        \item keywords: finger-worn, finger-based, finger-mounted, finger-attached, ring AND [Full Text] wearable (Publication Topics: Wearable Devices)
        \item applied filters: Conferences, Journals
    \end{itemize} 
}

To further narrow down the papers to those that fall within the scope of our survey, we then studied the abstract, introduction, and outline of each paper to decide whether we include the paper in the survey or not. Based on a series of inclusion and exclusion critera, two authors reviewed the 593 papers of our initial set and confirmed 159 papers to be selected for inclusion in our analysis (97 papers from ACM DL and 62 papers from IEEE). The inclusion and exclusion criteria were defined as follows:

\begin{enumerate}
    \item The paper has passed a peer review process.
    \item The work involves a ring form-factor device that follows our definition and ~\cite{shilkrot2015digital}'s ring form-factor. 
    \item We exclude ring-like devices that are worn on the entire hand, to differentiate between the concepts of rings and gloves, which is not defined by ~\citet{shilkrot2015digital}.
    \item We exclude papers that are not written in English.
    \item We exclude papers with no evaluation conducted.
    \item We exclude duplicate papers or those without an available PDF file.
\end{enumerate}

\begin{figure}
    \centering
    \includegraphics[width=1.0\linewidth]{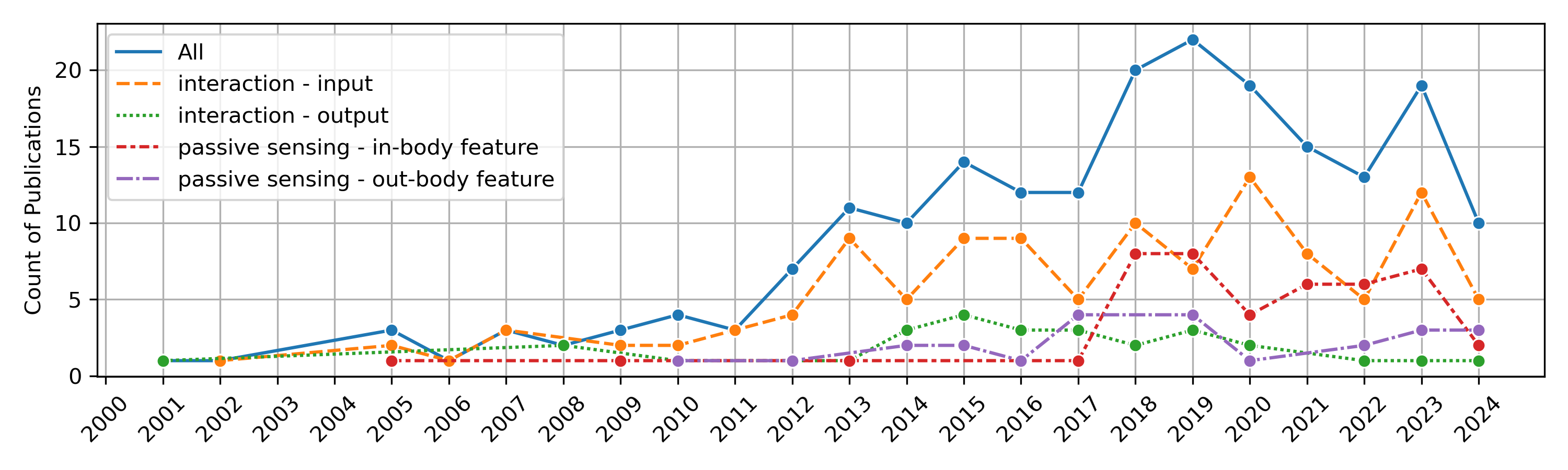}
    \caption{Annual publication count of smart ring research (total and per research area) from 2000 to 2024. One paper in the ``interaction - input'' category was published in 1997. }
    \label{fig:method_trend}
\end{figure}

Then, we performed backward-chaining to include papers that were published in other databases or missed by keyword-based querying. Papers were identified by reading the related work sections of the papers from the ACM and IEEE set and another author later confirmed the selection. This process resulted in 47 additional papers. Thus, our final set includes 206 papers in total.  Figure~\ref{fig:method_trend} shows the number of papers published per year since 2000. It shows an increasing trend of the research field with significant growth since 2013 and a peak in 2019.

To analyze and extract contents from these papers, we created a \textit{Google Sheets} table with 63 columns. The columns and the main taxonomy (the same as the outline from Section ~\ref{sec:interaction_input} to Section ~\ref{sec:passive_outbody}) were developed in an iterative manner. First, the two authors who scanned all papers for filtering purposes defined preliminary categories. This resulted in an initial list of 27 columns for all the papers to fill in. Then, two authors went through 15 papers from each category to further refine the main taxonomy and discuss the columns for each category to fill in. 
The papers were then split among authors for in-depth analysis.
The columns and main taxonomy were further refined as the reading progress proceeded.

In the primary taxonomy, we categorize the papers based on the users' intentions for using the smart ring: active interaction with the device, and passive usage where the ring can sense information without disrupting their current workflow. For the interaction category, we further divided it into input technology, where the user employs the smart ring as a specific input method, and output technology, where the smart ring delivers information to the user. The "interaction-input" category contains most of the smart ring publications, as shown in application overview Figure~\ref{fig:method_trend}. For passive sensing, we categorized the sensed phenomena into in-body features, predominantly physiological sensing, and out-body features, which include activity and environmental cues. These two categories contain fewer papers, possibly because smart rings for physiological sensing and activity recognition are already popular commercial products, such as the Oura Ring. Therefore, we also summarize the currently available health tracking commercial rings in Section~\ref{sec:passive_inbody}.
    
\section{Study Overview - Taxonomy} \label{sec:taxonomy}
Inspired by~\citet{roddiger2022sensing}, we adopted the following four-layer structure for the smart ring taxonomy. 

$$\textbf{Application} \rightarrow \textbf{Phenomena} \rightarrow \textbf{Fundamental Phenomena (FP)} \rightarrow \textbf{Sensors}$$
\vspace{-0.5mm}

\noindent We first describe the top-down categories regarding \textit{applications} in Section~\ref{sec:taxo_application}, with each category corresponding to one section in this paper. 
Then, we describe the \textit{phenomena} that enable these smart ring applications and the \textit{sensors} to detect them in Section~\ref{sec:taxo_phenomena}. 
Finally, inspired by ~\citet{vatavu2021gesturing}, we summarize smart ring wearing positions in Section~\ref{sec:taxo_position}.

\subsection{Application} \label{sec:taxo_application}

We adopted a \textit{``binary decision tree''} taxonomy for categorizing smart ring applications, where each node uses a classification rule to split it into two child nodes. At the root node, representing all smart ring literature in our identified set, we examined users' intentions for using the smart ring application. If the user actively uses the smart ring for interaction purposes, we categorize it under the \textit{``interaction''} category. Conversely, if the user does not actively interact with or use the smart ring, allowing the ring to passively sense information without disrupting their current workflow, we categorize this as the \textit{``passive sensing''} category. 

For the \textit{``interaction''} node, we further separate it into input and output technologies, which is a common taxonomy for interactive processes. In the \textit{``interaction - input''} (Section~\ref{sec:interaction_input}) category, the literature typically focuses on developing smart rings as substitutes for traditional input devices, such as keyboards or mice, or designing input interfaces on the ring itself. This category contains the majority of smart ring publications. Most literature from the \textit{``interaction - output''} (Section~\ref{sec:interaction_output}) category is concerned with designing haptic feedback, which is a common output channel for compact wearable devices~\cite{roddiger2022sensing, shilkrot2015digital}.

For the \textit{``passive sensing''} node, we further differentiate based on whether the sensed phenomena occur inside or outside the user's body. In the \textit{``passive sensing - in-body feature''} (Section \ref{sec:passive_inbody}) category, literature focuses on physiological parameters that can be measured on the hand or health state-related phenomena that can be derived from physiological signs. This category includes applications that monitor heart rate, blood oxygen levels, or other biometrics that are intrinsic to the user's body. In contrast, the literature in \textit{``passive sensing - out-body feature''} (Section~\ref{sec:passive_outbody}) category includes applications that analyze human activities and environmental cues observable by bystanders. Notably, different from ~\cite{roddiger2022sensing} where authentication is considered an independent category, we categorize authentication publications based on the type of feature used to identify user identity. For instance, an authentication method that utilizes the composition of a user's finger would be classified under the \textit{``passive sensing - in-body feature''} category whereas an algorithm that relies on the trajectory of a user's signature movement would fall under the \textit{``passive sensing - out-body feature''} category.

\begin{figure}
    \centering
    \includegraphics[width=1.0\linewidth]{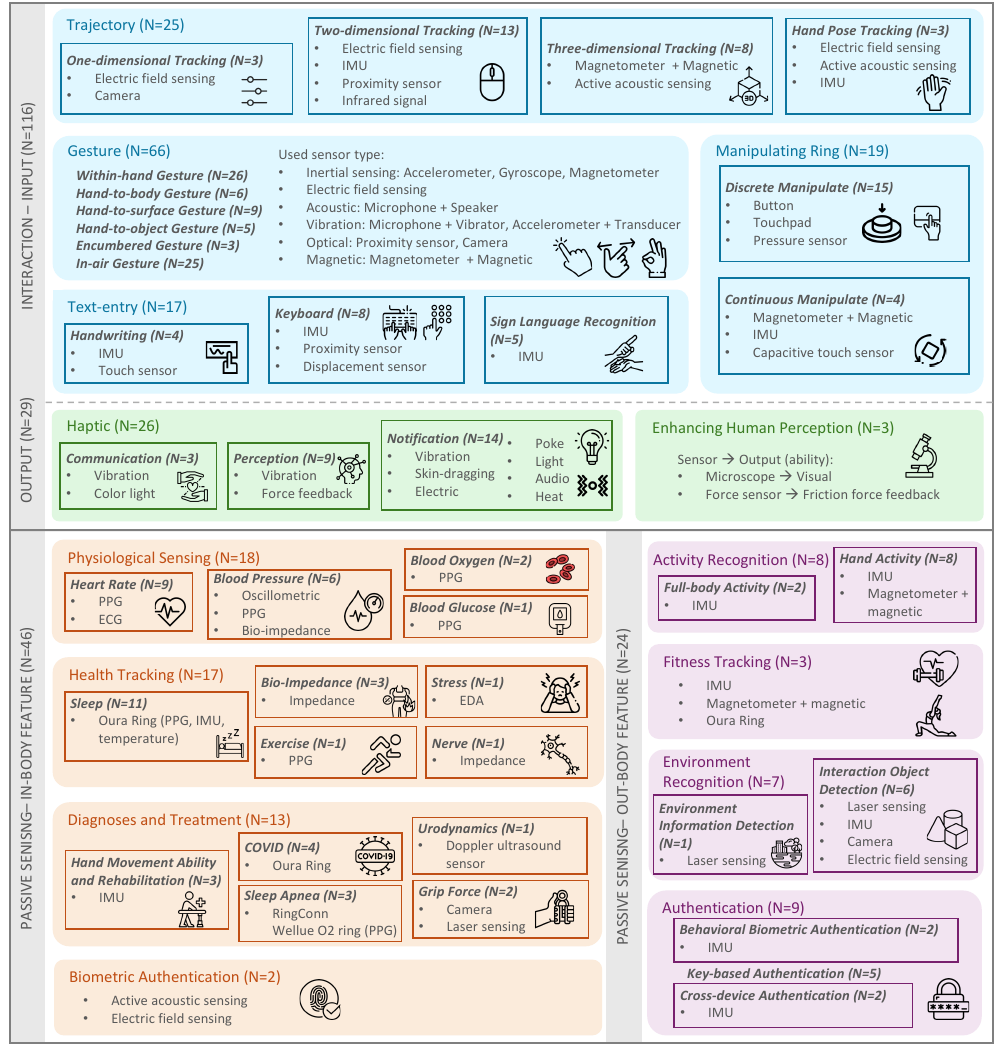}
    \caption{Taxonomy and Review Outline.}
    \label{fig:taxo_main}
\end{figure}

Figure~\ref{fig:taxo_main} provides an overview of the main structure of this survey, summarizing each category of applications identified in separate sections. For each subsection and subsubsection, we review the sensors, actuators, or commercial rings employed, as well as the number of papers associated with each category.

\subsection{Sensors and Phenomena} \label{sec:taxo_phenomena}

The majority of literature in Section~\ref{sec:interaction_output} relies on actuators, which can not fit in phenomena-based taxonomy, thus it is not considered in the taxonomy described in this subsection.

We first review the definition of \textbf{phenomena} as described in~\citet{roddiger2022sensing}. They define phenomena in the context of sensing-based applications as observable elements that are essential for structuring research work. This definition strategically bridges the gap between sensors, which serve as the tools for observation, and the applications, which are the ultimate targets of these observations. The term ``phenomena'' inclusively covers observations ranging from physiological parameters to broader conditions, events, states, or activities. Moreover, \textbf{fundamental phenomena (FP)} is defined as ``phenomena that can be captured directly by a sensor'' and ``can allow observation of other phenomena''.

To identify fundamental phenomena (FPs) for smart rings, we begin by summarizing sensor setups used in the literature. Table~\ref{tab:taxo_sensor} provides a description of sensor setups and the corresponding FP for each sensor. When determining the FP for each sensor, we examined the applications and implementations described in publications that utilize this sensor setup in their ring sensing solutions. The identified FP should represent the basic phenomena that all works in this set rely on. For instance, electric field sensing, which can be employed to measure body composition or predict hand gestures, fundamentally relies on the disturbance of the electric field caused by variations in body resistance. Thus, body resistance serves as the FP that underpins the various applications of electric field sensing in smart rings.

\begin{figure}
    \centering
    \includegraphics[width=1.0\linewidth]{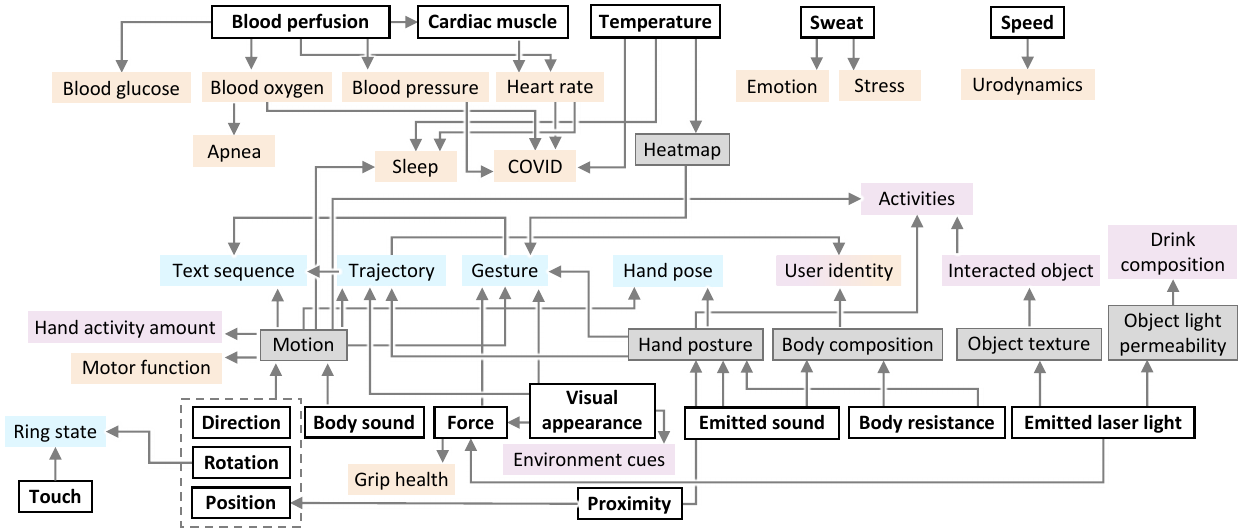}
    \caption{Flow diagram showing how different phenomena can be inferred. The white boxes represent “fundamental” phenomena that can be directly sensed and from which all other phenomena can be derived.}
    \label{fig:phenomena_flow}
\end{figure}

Based on the FPs we identified, we proceeded to determine the additional phenomena that can be derived from these FPs, resulting in the final set of phenomena for smart ring sensing. Figure ~\ref{fig:phenomena_flow} presents a flow diagram illustrating the progression from ``$Phenomena \rightarrow FP$''. In this diagram, FPs are depicted in white boxes, while other boxes represent non-fundamental phenomena. Using this flow diagram, we are able to trace the ``$Application \rightarrow Phenomena \rightarrow Fundational Phenomena(FPs) \rightarrow Sensors$'' pathways for publications included in our literature review (except for Section~\ref{sec:interaction_output}). For the last phenomena node in each route, the boxes are colored to match the application category colors used in Figure~\ref{fig:taxo_main}.

\begin{table*}[htbp]
    \centering
    \footnotesize
    \caption{Sensor integrated in smart rings and fundamental phenomena that can be observed by the corresponding sensor.}
    \resizebox{1.0\textwidth}{!}{%
    \begin{tabular}{p{3cm}p{9cm}c}
    \hline
    \textbf{Sensor} & \textbf{Description} & \textbf{FP} \\
    \hline
    Accelerometer & Accelerometer directly measures proper acceleration relative to a free-falling observer. It can be used to infer static information like pointing direction or dynamic hand movements. & direction \\
    \hdashline
    Gyroscope & Gyroscope measures the orientation and angular velocity, which indicates hand motion. & rotation \\ 
    \hdashline
    Magnetometer & Magnetic field changes caused by hand movements or body activity. &  (absolute) direction  \\ 
    \hdashline
    Magnetic + Magnetometer & Typically used in the form of a magnetic ring and another device contains a magnetometer. This setup allows the magnetometer to detect changes in the magnetic field caused by the movements and positioning of the ring. & (relevant) position \\
    \hline

    Microphone & Utilized as contact microphones to capture the subtle sounds produced by hand motions or the frictional sounds emitted when fingers interact with various objects. & body sound \\
    \hdashline
    Active acoustic sensing & Microphone capture the sound waves that travel through the hand or reflect off it, influenced by factors such as the hand pose, the biological features of the hand, etc. & emitted sound \\
    \hdashline
    Continuous Wave Doppler Ultrasound Sensor & Measure the reflected frequency of ultrasound, sensitive to the motion of tissues or fluids. & speed \\
    
    \hline
    Photoplethysmography (PPG) & Emitting light into the skin and measures the amount of reflected light. The amount of light absorbed varies with the pulsation of blood vessels.  & blood perfusion \\
    \hdashline
    Infrared Thermometry & Non-contact temperature measurement by detecting the intensity of infrared radiation. & temperature\\
    \hdashline
    Proximity Sensor & Detect the presence and distance of close objects based on the intensity of the reflected infrared light. & proximity\\
    \hdashline
    Laser Sensing  & Use CCD sensor or photo-resistor to sense the properties of the reflected laser light, which can be used to recognize texture or object deformation. & emitted laser light \\
    \hdashline
    Camera & Capture hand pose or environmental cues. & visual appearance \\

    \hline
    Electrodermal Activity (EDA) & Measures skin's electrical conductance that varies with its moisture level, which can be used to assess emotional responses and stress levels. & sweat \\
    \hdashline
    Impedance/Electric field sensing & Operate by emitting a small electric field and measuring disturbances caused by the presence of conductive or dielectric objects within this field. This allows the sensor to detect hand contact, position, physiological parameters such as body composition, etc. & body resistance \\
    \hdashline
    Electrocardiography (ECG) &  Measures the electrical signals associated with cardiac activity. & cardiac muscle \\
    \hline
    
    Button & Binary state indicator pressed to use certain functions. & touch \\
    \hdashline
    Touchpad & Detect the motion and position of a contact finger. & touch \\
    \hdashline
    Pressure Sensor & Force level. & force \\
    \hdashline
    Thermistor & Temperature indicated by resistance change. & temperature \\
    \hline
    \end{tabular}

    }
    
    \label{tab:taxo_sensor}
\end{table*}

Table~\ref{tab:taxo_phenomena} provides the description for all identified phenomena.

\begin{table*}[htbp]
    \centering
      \footnotesize
      \caption{Definition of all phenomena investigated with smart rings. Boldface denotes fundamental phenomena (FP) that enable observation of higher-level phenomena.}
    \resizebox{1.0\textwidth}{!}{%
    \begin{tabular}{p{3cm}p{14cm}}
    \hline
    \textbf{Phenomena} & \textbf{Description} \\
    \hline
    \textbf{Direction} & Geometrical orientation relative to earth coordinate.\\
    \textbf{Rotation} & Circular movement around axis. \\
    \textbf{Position} & The location of the ring relative to the object. \\
    \textbf{Touch} & User touches the ring itself for interaction. Touch gestures performed on other surfaces or with-in hand are considered as ``Gesture''. \\
    \textbf{Body sound} & Sound emitted by the wearer or inside the wearer's body. Hand movement can produce sound such as friction sound~\cite{zhang2017fingorbits} or tapping sound that can propagate through bone conduction~\cite{fukumoto2005finger}. \\
    \textbf{Proximity} & A short distance measured from the ring to objects or fingers. Due to the dexterous and close nature of fingers, proximity can be used to infer hand posture~\cite{sun2021thumbtrak, kienzle2014lightring}.\\
    \textbf{Force} & The strength that hand exerts to object. Exerting force usually can cause fingernails to turn white~\cite{urban2013computing, li2022nailring} or cause objects to deform~\cite{yin2023hippo}.\\
    \textbf{Visual appearance} & Capture surroundings' image that can reveal context information. \\
    \textbf{Emitted sound} & Sound emitted by the active acoustic system on the ring and then captured by the microphone. The difference of reflected sound may indicate hand posture~\cite{yu2024ring} or the sound propagated through the finger may indicate body composition~\cite{bianchi2017disambiguating}.\\
    \textbf{Body resistance} & Natural opposition offered by hand to the flow of electric current, which is influenced by factors such as skin condition, body composition~\cite{usman2019analyzing}, hand posture~\cite{waghmare2023zring}, etc. \\
    \textbf{Emitted laser light} & Laser light emitted by the laser sensing system on the ring and then sensed. The sensed laser light can be relevant to the refraction~\cite{yin2023hedgehog} or reflection~\cite{wang2024texturesight} process. \\
    \textbf{Blood perfusion} & The rate at which blood is delivered to tissue.\\
    \textbf{Cardiac muscle} & Rapid, involuntary contraction and relaxation of the cardiac muscle for pumping blood throughout the cardiovascular system.\\
    \textbf{Temperature} & Measure body temperature at hand or environment temperature to form a heatmap.\\
    \textbf{Sweat} & Fluid secreted by sweat glands in the skin. \\
    \textbf{Speed} & The speed of certain objects or liquids.\\
    \hline
    Motion & The change of position and direction in a period of time. The dexterous nature of the hand provides rich motion information.\\
    Hand posture & General orientation and position of the hand and fingers.
    \\
    Object texture & Tactile and visual properties of an object's surface, including its roughness, smoothness, softness, or pattern.\\
    Light permeability & The ability of an object to allow light to pass through it, depending on the material and thickness.\\
    Heatmap & The distribution and variation of temperatures across a specific area or surface.\\
    \hline
    Ring state & The physical attributes of the software are monitored by its computing chip.\\
    Gesture & A physical movement or signal performed by the body. \\
    Trajectory & The path that a finger moves through space in a period of time. \\ 
    Text sequence & Hand motion are performed to convey an ordered series of words, characters, or symbols. \\
    Hand pose & Detailed configuration and arrangement of each finger and the hand itself, normally represented in 21-DoF hand pose~\cite{zhou2023one} (see Figure~\ref{fig:taxo_hand_dof}). \\
    \hdashline
    Blood glucose & Concentration of sugar in the bloodstream. \\
    Blood oxygen & The percentage of oxygen-saturated hemoglobin in the blood. \\
    Blood pressure & The force exerted by circulating blood on the walls of blood vessels\\
    Heart rate & The number of heartbeats per minute\\
    Apnea & Temporary cessation of breathing during sleep. \\
    Sleep & Natural, cyclic state of rest for the mind and body. Smart rings usually detect the stage of sleep or the change between awake and asleep.\\ 
    COVID & Detect the COVID-19 infectious respiratory disease. \\
    Emotion & Psychological state that involves an individual's subjective experience, physiological response, and behavioral expression. \\
    Stress & The body's response to any demand or challenge, which can be physical, mental, or emotional. \\
    Urodynamics & Assess the function and efficiency of the bladder and urethra in storing and releasing urine. \\
    Motor function & The ability of the nervous system to coordinate muscle activity for hand movement. \\
    Grip health & The strength and functionality of the hand muscles. \\
    Body Composition & The percentages of fat, muscle, bone, and water in the human body. \\
    \hdashline
    Hand activity amounts & Quantifies the frequency and intensity of hand movements. \\
    Environment cues & Signals or indicators within a surrounding space, without contact with users.\\
    Activities & Physical and mental tasks or engagements that an individual performs. \\
    Interacted objects & Objects that the user physically engages with during daily activities\\
    Drink Composition & The ingredients in a beverage, related to personal safety~\cite{yin2023hedgehog}.\\
    \hdashline
    User identity & The unique set of characteristics that distinguish one user from another. \\
    \hline
    \end{tabular}
    }
    \label{tab:taxo_phenomena}
\end{table*}

\begin{figure}
  \centering
  \subcaptionbox{Hand Skeleton \label{fig:taxo_hand_skeleton}}
    {\includegraphics[width=0.35\linewidth]{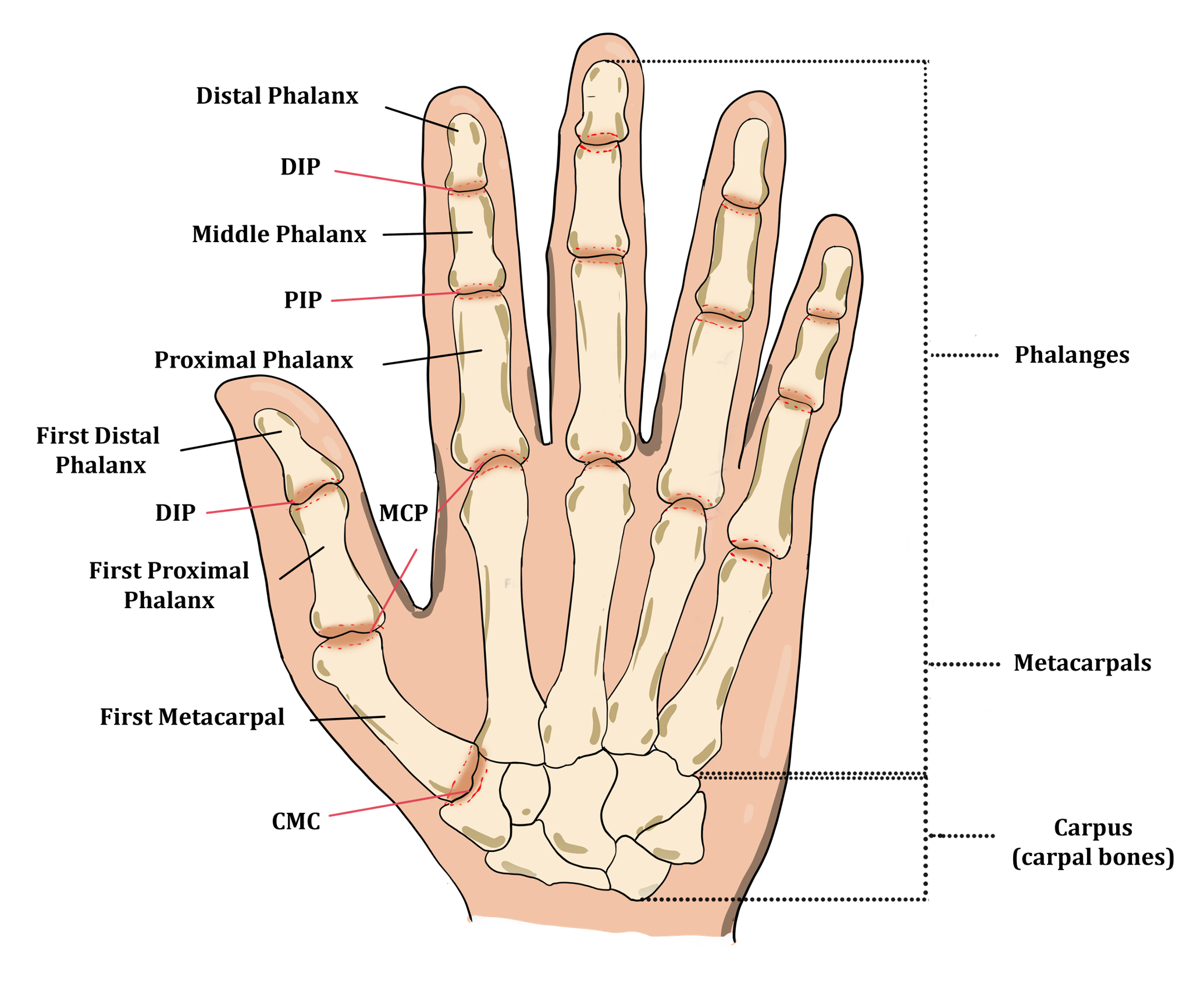}}
  \subcaptionbox{Joint DOF\label{fig:taxo_hand_dof}}
    {\includegraphics[width=0.35\linewidth]{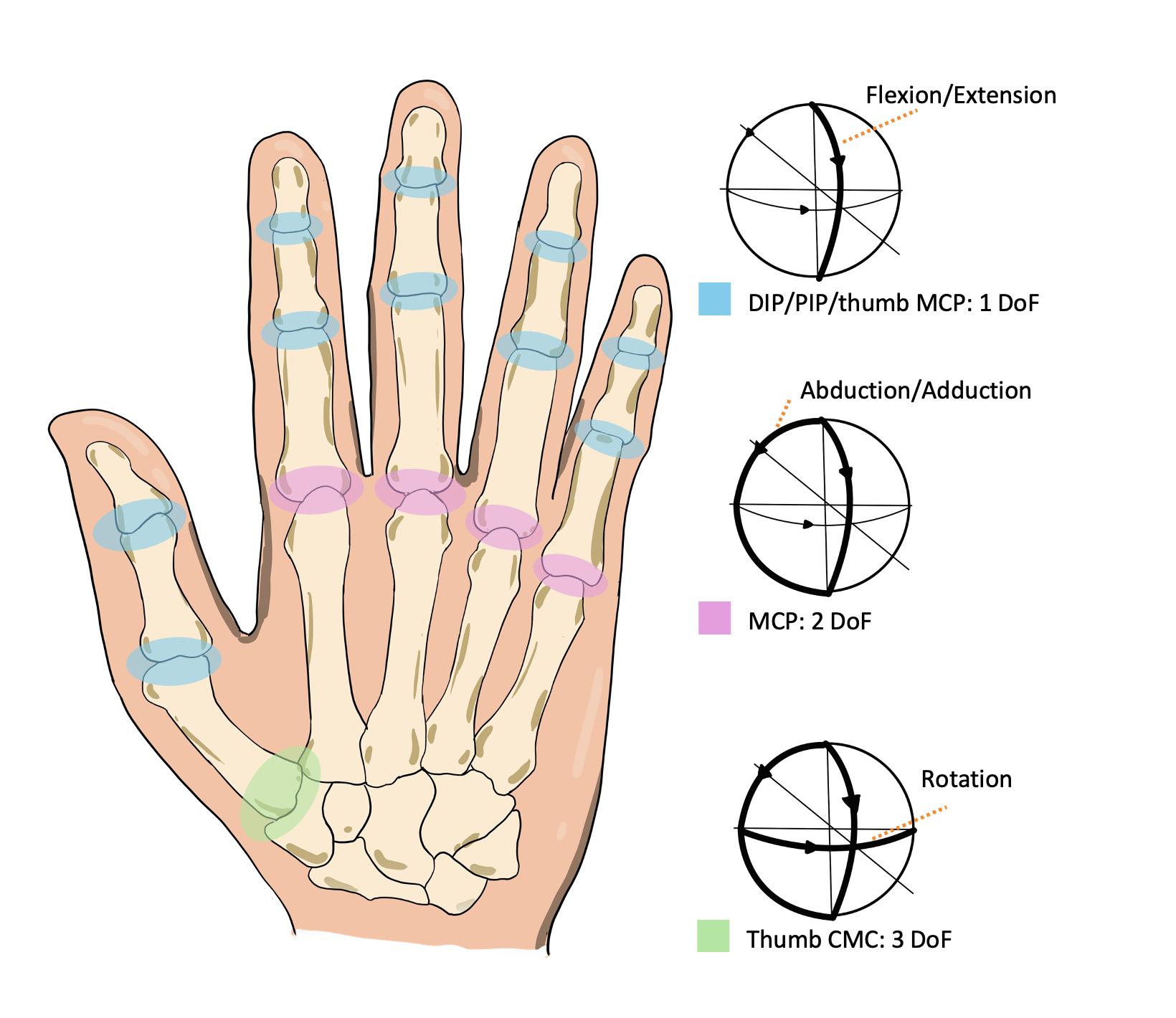}}
   \subcaptionbox{Wearing Position\label{fig:taxo_position}}
    {\includegraphics[width=0.25\linewidth]{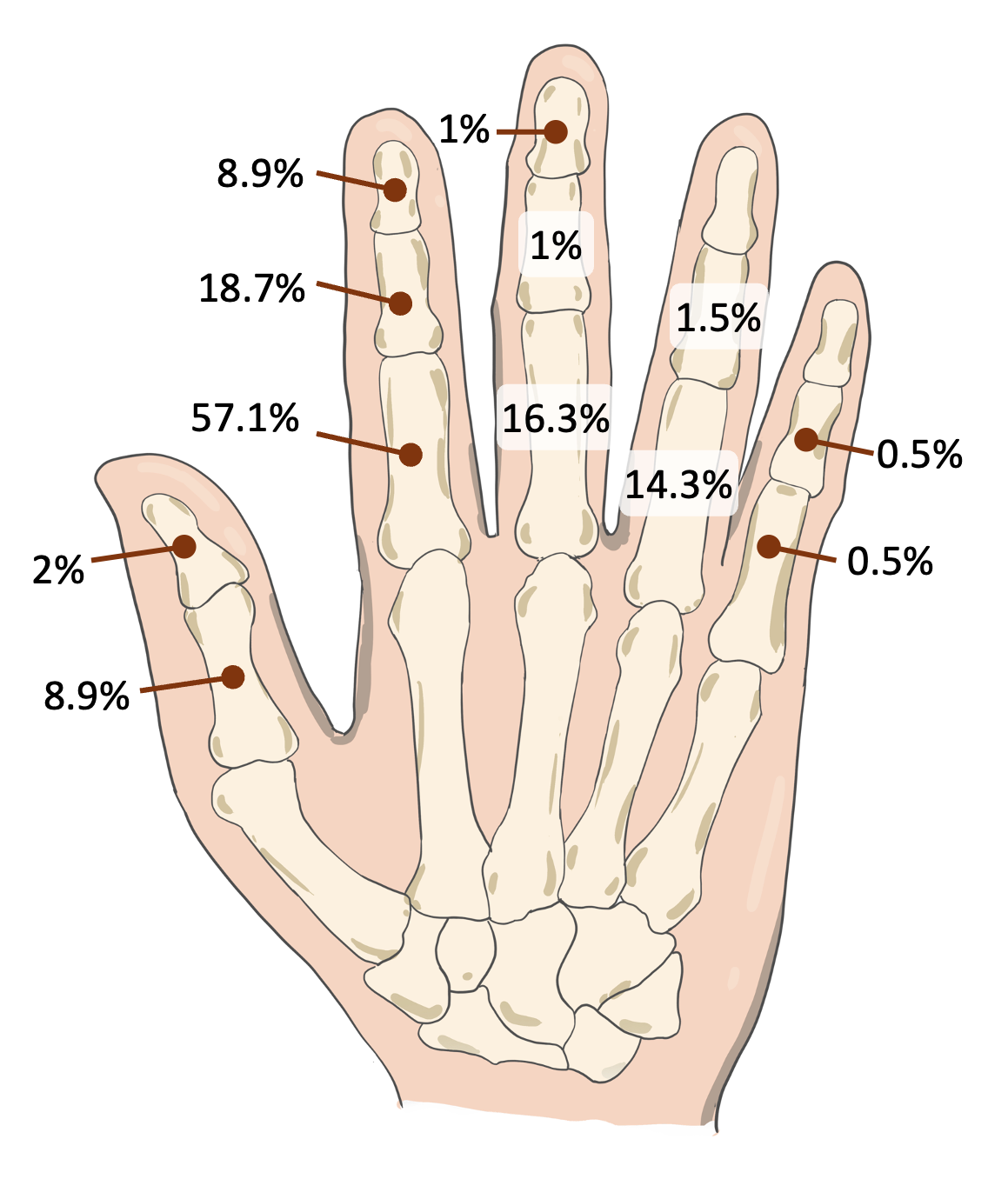}} 
  \caption{Hand anatomy (a), degrees of freedom (b) and unifying terminology for the wearing position of rings (c).}
  \label{fig:taxo_hand_terminology}
\end{figure}

\subsection{Hand Terminology and Ring Wearing Position} \label{sec:taxo_position}

Figure~\ref{fig:taxo_hand_skeleton} illustrates the terminology for the hand skeleton that is utilized throughout this survey paper and the literature we have reviewed. The flexibility of fingers enables a wide range of interaction applications with smart rings. Figure~\ref{fig:taxo_hand_dof} demonstrates the degrees of freedom (DoF) for each finger joint, totaling 21 DoF for each hand. Figure~\ref{fig:taxo_position} shows the frequency of smart rings being worn on different phalanges. The publications from which the unifying terminology and statistical results are derived exclude three papers in which the specific phalanx for ring placement is neither mentioned nor can be inferred. The percentages illustrated in Figure~\ref{fig:taxo_position} do not sum to 100\% because some studies require smart rings to be worn on multiple phalanges~\cite{liang2021dualring, shen2024mousering, nicolau2013ubibraille} or allow for the smart ring to be worn on various phalanges~\cite{xu2022washring, rajput2023assessment, roumen2015notiring}.

\section{Interaction - input} \label{sec:interaction_input}

\begin{figure}
    \centering
    \includegraphics[width=0.9\linewidth]{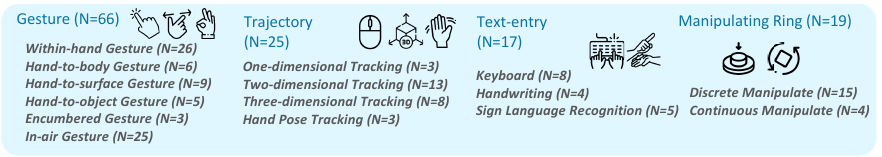}
    \caption{Section outline for interaction input.}
    \label{fig:input_outline}
\end{figure}

The dexterous nature of hands offers a broad design space for interaction input, which is particularly advantageous in the context of smart rings. These devices can be utilized for a variety of input methods, including gesture-based inputs, using the smart ring to predict input signals for other virtual input devices, or directly manipulating the ring itself as an input interface. The following of this section is structured as Figure~\ref{fig:input_outline}. Section~\ref{sec:interaction_input_mani} details the literature involving actions performed directly on the ring, such as using input interfaces like buttons or altering the ring's position and/or orientation relative to the finger. The other three subsections focus on literature involving actions performed with the ring, where the ring's position and orientation relative to the finger remain unchanged during the application's use. Specifically, Section~\ref{sec:interaction_input_gesture} covers literature related to gesture recognition, Section~\ref{sec:interaction_input_traj} discusses continuous trajectory recognition, and Section~\ref{sec:interaction_input_text} explores implementations of text-entry using the ring.

\subsection{Gesture} \label{sec:interaction_input_gesture}

\begin{figure}
    \centering
    
    \includegraphics[width=0.7\linewidth]{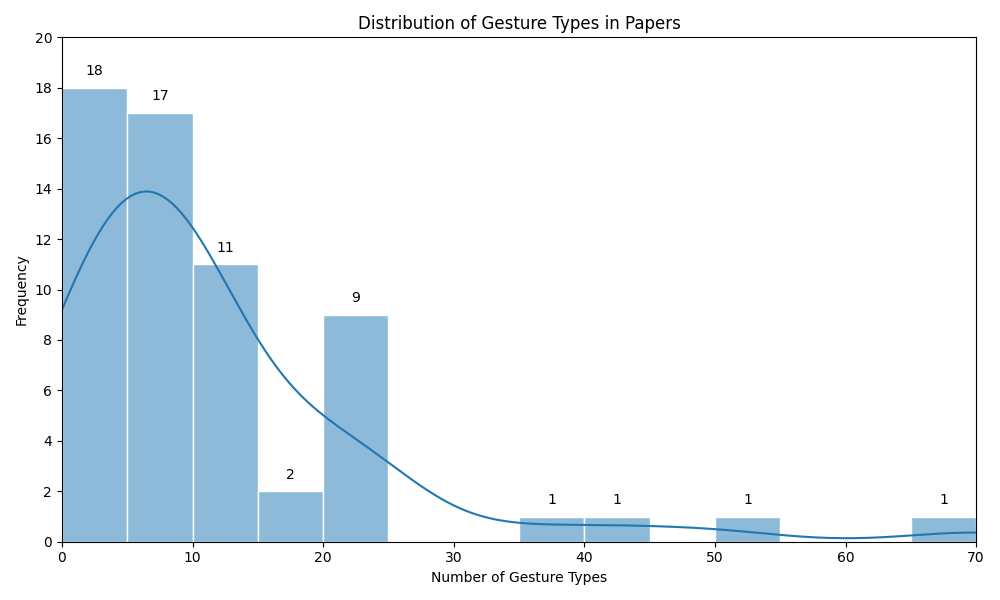}
    \caption{Distribution of Gesture Set Scale by Type.}
    \label{fig:histogram_gesture_type}
\end{figure}

As naturally dexterous tools, hands serve as an exceptionally intuitive medium for human-computer interaction. Consequently, the use of rings worn on the hands to perceive and interpret gestures for the purpose of completing interactive input tasks has garnered considerable attention from researchers. Owing to the privileged wearing position of smart rings, they are endowed with the ability to sense various gestures including coarse-grained and fine-grained, which is much more suitable than a smartwatch or wristband. We first analyzed gesture types explored in papers and provided the distribution in Figure~\ref{fig:histogram_gesture_type}. Inspired by the idea of ~\cite{liang2021dualring}, we build upon the taxonomy of different gestures progressively from fine-grained to coarse-grained as 1) within-hand gesture, 2) hand-to-body gesture, 3) hand-to-surface gesture, 4) hand-to-object gesture, 5) encumbered gesture, and 6) in-air gesture. We also provided high-frequency types of gestures as shown in Table~\ref{tab:gesture_distribution}. In the following part of this section, we will introduce each category in detail.

\begin{table*}[!t]
    \centering
    \caption{Number of publications investigating specific hand gesture types, including gesture types that appear more than once.}
    \resizebox{0.7\textwidth}{!}{%
    \begin{tabular}{m{4cm}m{3cm}m{4cm}p{2cm}}
    \hline
    \textbf{Major Categories} & \textbf{Subcategories} & \textbf{Variants} &\textbf{Frequency}\\
    \hline

    \multirow{7}{*}{Within-hand Gesture} & Pinch & - & 7 \\
    \cdashline{2-4}
     & Touch 12 knuckles & - & 6 \\ 
    \cdashline{2-4}
     & Draw path & Triangle/Circle/Letter/Digit \newline On finger/In palm & 4 \\ 
    \cdashline{2-4}
     & Swipe & Left/Right & 9 \\
    \cdashline{2-4}
     & Tap & Finger/Palm/Hand back & 10 \\
    \cdashline{2-4}
     & Long tap & - & 2 \\
    \cdashline{2-4}
     & Double tap & - & 2 \\
    \hline


    \multirow{2}{*}{Hand-to-surface Gesture} & Tap & - & 12 \\
    \cdashline{2-4}
     & Swipe & Finger pad/Fingertip  & 7 \\ 
    \hline


    \multirow{2}{*}{Encumbered Gesture} & Swipe & - & 2 \\
    \cdashline{2-4}
     & Tap & - & 2 \\ 
    \hline

    \multirow{4}{*}{In-air Gesture} & Point & 6 directions & 9 \\
    \cdashline{2-4}
     & Rotate & Clockwise/Anticlockwise & 4 \\ 
    \cdashline{2-4}
     & Draw path & Circle/Square/Digit/Letter & 11 \\ 
    \cdashline{2-4}
     & Sign language & Digit/Letter/Word/Sentence & 5 \\
    \hline

    \end{tabular}
    }

    \label{tab:gesture_distribution}
\end{table*}

\subsubsection{Within-hand Gesture}

Within-hand gesture mainly incorporates gestures of single-hand gestures and double-hand gestures, where the former includes interactions between the thumb and the other four fingers and the latter includes interactions between the left hand and the right hand. 

These gestures include pinch, tap, swipe, and various subcategories that originate from them such as double tap and long tap. ~\citet{fukumoto2005finger} utilized the vibration generator and microphone embedded in the ring to detect pinch and achieve the control of audio, which was an early test of pinch. As a follow, Tickle~\cite{wolf2013tickle} used more commonly used IMU to detect pinch, achieving the control of handheld devices. In 2015, Cyclopsring~\cite{chan2015cyclopsring} creatively used the fisheye camera to detect pinch in order to interact with the environment object. Making the detection more accurate, Electroring~\cite{kienzle2021electroring} and Dualring~\cite{liang2021dualring} developed the technical of using electric-field and high-frequency AC circuit to identify the timing of pinch. 

Apart from the classical pinch, lots of work has been done to support different within-hand gestures. One of the most common designs is using a single hand to perform microgestures, which indicates the gestures between the thumb and the other four fingers. Dualring~\cite{liang2021dualring} utilized the index finger as the interface to conduct swipe and tap, which is also commonly seen in other works. They also implemented the popular gesture of touching 12 knuckles as items and drawing specific paths on the finger. In EFRing~\cite{chen2023efring}, electric-field was designed to detect different microgestures including tap, double tap, swipe of four directions and draw tick, clockwise loop, and anticlockwise loop. Touchring~\cite{tsai2016touchring} also introduced electrode capacitive sensors to perceive the tap and swipe of thumb on the index in cases of middle finger touch or non-touch the index. Apart from electric sensors, proximity light is also seen as accurate in detecting the time of single-hand microgesture. ThumbTrak~\cite{sun2021thumbtrak} adopted the proximity sensor to sense the touch on 12 knuckles. Similarly, ~\citet{kokubu2022one} took photoreflectors as proximity to identify swipe and drawing circles. OptiRing~\cite{waghmare2023optiring} implemented the miniature endoscopic camera and LEDs on the ring to detect the microgestures between the thumb and index. To detect the timing directly, Thumb-in-motion~\cite{boldu2018thumb} directly implemented the touchpad on the ring and figured out the swipe and tap when standing, walking, and running. Like a touchpad, ~\citet{ghosh2016ringteraction} used the capacitive touch sensors to detect tapping, swiping, and drawing circles on the index finger with the thumb. Using capacitive sensing, PeriSense~\cite{wilhelm2020perisense} detected the distance of different fingers, which can be used to identify the bending of the hand. To make a better understanding of the combination gesture of swiping and scrolling on the index finger, ~\citet{herath2022expanding} implemented a capacitive sensor on the ring and achieved the classification of the gestures.

Compared with other sensors, the most commonly used sensors are IMUs and microphones. FingerSound~\cite{zhang2017fingersound} employed the combination of IMU and microphone to figure out the drawing path of the thumb on the palm, achieving the identification of 42 different within-hand gestures including swipe, hand-write numbers, and hand-write letters. FingerPing~\cite{zhang2018fingerping} made use of the speaker and microphone to detect the tap on 12 knuckles. Similarly, ThumbRing~\cite{tsai2016thumbring} took the 9-Dof IMU to collect motion information of the thumb tapping on the 12 knuckles and bound them to different items of the devices. FingOrbit~\cite{zhang2017fingorbits} also combined the microphone and IMU to distinguish the friction of the thumb on the other fingers at different speeds. Vibaware~\cite{kim2023vibaware} took up the vibrator and the accelerometers to discriminate the tap and swipe in hand, on the surface, and when grasping. By combining the capacitive touch sensor and the IMU, ~\citet{meli2017hand} achieved the text entry by tapping the knuckles. With the help of a 3-axis accelerometer, ~\citet{iwamoto2007finger} implemented the measure of vibration on the knuckles to distinguish different taps.

To make the ring energy consumption lower, lots of designs have also been implemented. The most commonly used design is a magnet so that there is less need to contain the battery in the ring. SynchroWatch~\cite{reyes2018synchrowatch} introduced a ring made of a permanent magnet and uses the magnetometer in the watch to detect the touch between thumb and finger whose frequency is designed to select different components in the watch. PicoRing~\cite{takahashi2024picoring} also used the permanent magnet as the ring, combined with which a wristband coil is able to identify the swiping, tapping, and scrolling mircogesture. 

In Z-Ring~\cite{waghmare2023zring} and MType~\cite{deng2019mtype}, double-hand gestures were studied. The former took the other hand back as an interface to conduct tap(single, double, and long) and swipe(left and right), which is detected by multi-frequency impedance sensing. The latter took the other hand back and palm as interface and achieved tap in different places for different items, which is identified by the combination of watch IMU and magnet ring.

\subsubsection{Hand-to-body Gesture}

The human body is of great area and provides abundant possibility of interface. Lots of gestures are designed to make full use of it. WashRing~\cite{xu2022washring} utilized a ring with IMU to recognize different gestures in the process of washing hands to support the nine-step handwashing method. ~\citet{li2023enabling} designed 15 types of hand-to-head gestures to stand for different voice interaction goals, which are identified by the IMU in the ring. ~\citet{liu2020keep} adopted taps and swipes conducted on the thigh to support the visually impaired users operating a smartphone, which is under the assistance of a ring with IMU. Using 3 rings worn on the hand, ~\citet{moschetti2016recognition} achieved the detection of 9 different gestures in 9 types of gestures used in the daily lives of elderly people including answering the phone and brushing their teeth. ~\citet{mujibiya2013sound} introduced the ultrasound and used the ring as a receiver to sense the swipe and grasp on the arm. Using infrared sensors, an IMU, and a mini camera, TouchCam~\cite{stearns2018touchcam} achieved click detection on 15 different parts of the body.

\subsubsection{Hand-to-surface Gesture}

It is common to perform gesture interaction on a surface, which is able to provide sufficient feedback and adequate area for different operations. Widely used hand-to-surface gestures include tap, swipe, scroll, and drawing specific path and their variant. In Dualring~\cite{liang2021dualring}, the tap on the surface is supported by the AC circuit. ~\citet{gu2019accurate} discussed the impact of the different fingers, finger positions, and hand status on the interaction of tap on the surface, which is achieved by a ring with IMU. Except for the pinch mentioned above, Tickle~\cite{wolf2013tickle} also compared the tap and swipe on different surfaces including flat, convex, and concave curved with an IMU ring mounted on the fingertip. Moreover, supported by two rings with IMUs Touch+Finger~\cite{lim2018touch+} designed 20 different gestures using the free finger when the index touches the screen, providing more space for interaction on the screen surface. Using the accelerometer, force-sensitive resistor, and audio sensor, ~\citet{gummeson2014energy} achieved the detection of swipe, scroll, and 12 basic strokes of letters on the device surface. Creatively, NailRing~\cite{li2022nailring} made use of the camera to perceive the subtle color change of the fingernail to detect swipes and taps on the surface and thumb. To detect the tap on the surface accurately, ~\citet{shi2020ready} utilized a single IMU to achieve the transfer of identification of tap and release on the surface of different rigidity. With the microphone, accelerometer, and gyroscope, ~\citet{zhang2011ring} captured the bone-conducted sound and finger movements on a surface to control the screen, supporting interactions of swipe and tap. \citet{han2016exploring} utilized an IMU sensor embedded in a smart ring to capture vibration signals generated when a finger swipes over a surface, allowing users to perform swipe gestures on a cutting board to interact with a touchscreen during cooking, or to execute swipe gestures on different parts of non-uniform textured objects to enable contextual actions.


\subsubsection{Hand-to-object Gesture}
Although hand-to-object gestures are rare in the research community, it is of value to discuss them. Similar to hand-to-surface gestures, these gestures need to touch the object directly to get enough feedback. Differently, the object in hand-to-object gestures always refers to those with a specific shape or type, rather than a common surface. Dualring~\cite{liang2021dualring} detected whether the user is gripping an object and turns the surface of the gripped object into an interactive interface to finish some control of the television or light. MagTouch~\cite{park2020magtouch} used a permanent magnet as the ring to distinguish the index, middle, and ring finger when tapping on the watch screen. Similarly, ~\citet{oh2021identifying} utilized speakers and microphones to discriminate the thumb, index, and middle finger when tapping on the tablet. ~\citet{bianchi2017disambiguating} used the vibration and accelerometer to distinguish the users and fingers. ~\citet{gupta2018asterisk} designed different one-stroke drawings on different objects, following which with the index finger, the objects can be identified like using a QR code.

\subsubsection{Encumbered Gesture}
One of the most important advantages of the ring in the interaction field is that the ring worn on the finger is always available, which means that it is possible to access the interface at any time, even when hands are encumbered by other objects. Different from the hand-to-object gestures, objects in encumbered gestures are always irrelevant to the interaction itself. In other words, hand-to-object gestures mean that the users want to interact so they hold the objects, while encumbered gestures mean that users want to interact in a situation where they have to hold something. 

ARO~\cite{bardot2021aro} discussed the zooming and panning gestures in ARO(in-air, on-ring, on-object) when the hand is encumbered, by grasping or holding objects and finds that users are most efficient when using either in-air or on-ring interactions and in-air is the most preferred by users, which is supported by an IMU and a touchpad. ~\citet{wolf2016microgesture} achieved the identification of the microgestures of swipe and tap when riding a bicycle and holding the handlers using two IMUs. ~\citet{wolf2013ubiquitous} placed two 9-DoF IMUs on the fingertip and designed 21 gestures to interact when grasping something.

\subsubsection{In-air Gesture}

Another widely used gesture design is the in-air gesture, which possesses the advantage of flexible posture and wider design space. Different from the microgesture and hand-to-surface gesture, lack of physical feedback leads to the missing of swipe and tap. Instead, the most common gestures are pointing and drawing a specific path. Apart from these, in-air gestures also include zoom, rotation, snap, sign language, and other actions combined with the arm. 

The basic subset of in-air gestures is fundamental gestures including pointing, rotating, swinging, and similar primary actions. ~\citet{mousavi2023wearable} designed 9 in-air gestures that were able to be distinguished by a ring with the IMU, including finger-pointing in different directions, hand rotating clockwise, and zooming in/out. Using a permanent magnet ring, ~\citet{ketabdar2010towards} designed 8 in-air gestures to interact with a smartphone, including approaching, away, and drawing a circle. Similarly, MagiTact~\cite{ketabdar2010towards} also used a permanent magnet ring to design in-air gestures to control the smartphone. Supported by the electro field sensors, eRing~\cite{wilhelm2015ering} was able to detect 12 in-air gestures, such as grasping, fist, and snap. To help visually impaired users, ~\citet{namdev2015interactive} introduced a nod ring with motion sensors to achieve the zooming operation using the vertical motion of hands. ~\citet{yamagishi2014system} used a similar device like MagicRing~\cite{jing2013magic} with an acceleration sensor to achieve 23 control gestures of the personal computer, such as push, pull, swing, and so on. More completely, MobiRing~\cite{chen2014mobiring} discussed all of the 6 directions of hand motion using an accelerometer. Utilizing four infrared sensors, iRing~\cite{ogata2012iring} could identify information such as skin deformation and pressure, and subsequently infer finger rotation, finger bending states, and gestures. Designing 6 types of air gestures including pointing and scanning, i-Throw~\cite{lee2007ubiquitous} enabled interaction with the proposed UFC wearable device. Like MagicRing~\cite{jing2013magic}, ~\citet{jing2011recognition} used an accelerometer to detect 12 types of one-stroke in-air gestures including rotating, finger crooking, and shifting. ~\citet{ketabdar2012pingu} presented a ring concept as an input device for around-device interaction with 9-DoF IMU and 2 proximity sensors for finger movement and orientation detection. They evaluated the proposed system's effectiveness with 9 commonly used in-air gestures and achieved 97\% classification accuracy.

Another important subset of the in-air gestures is drawing a specific path in the air. ~\citet{cheng2018finger} used the 3-DoF IMU to capture the motion of the hand to classify 24 different paths, including squares, circles, and so on. TinyDL~\cite{coffen2021tinydl} collected the real-time IMU data to detect paths of drawing the nine Arabic numerals and zero. Similarly, ~\citet{zhou2013adaptive} also used the MagicRing to distinguish 8 drawing paths. Combining the smartwatch and the ring with IMU, WRIST~\cite{yeo2019wrist} achieved the interaction of drawing circles, squares, triangles, and question marks. Unlike the former work, ThermalRing~\cite{zhang2020thermalring} introduced a thermo sensor to the ring, based on several different paths of drawing letters. Using the haptic feedback, ~\citet{yeom2015poster} achieved the identification of Korean characters, uppercase and lowercase English letters, and numbers written in the air. RingIoT~\cite{darbar2019ringiot} designed 7 types of in-air drawing paths to control the IoT devices using a ring combined with an IMU, a capacitive sensor, and an IR emitter. To support the in-air drawing of custom path, ~\citet{xie2015similarity} designed 8 basic gesture paths (8 directions) and derived 12 complex gestures based on them where the complex gestures are decomposed and detected using a smart ring supported by a 3-axis IMU.

As for the sign language, FingerPing~\cite{zhang2018fingerping} utilized the microphone and speaker to detect digits ’1’ to ’10’ from American sign language. MagicRing~\cite{jing2013magic} implemented the perception of pointing up/down, rotating left/right, and 8 hand talk gestures using a ring with IMU. ~\citet{liu2020finger} combined the IMU data from the ring and watch to detect hand poses for alphabets. ~\citet{kuroki2015remote} focused on the simplified Japanese sign language and achieved the discrimination of 10 different hand talk gestures using the ring with IMU. ~\citet{liu2021video} combined the IMU and open video from the video to train a model, which achieves the distinction of 50 gestures from the American sign language tutorial.

\subsection{Trajectory} \label{sec:interaction_input_traj}

\subsubsection{One-dimensional Tracking}
Detecting one-dimensional (1D) motion trajectories—such as a finger sliding along another finger or a linear surface—significantly advances interaction paradigms prioritizing precision, convenience, and discretion.

This capability can be achieved through electric field sensing techniques. EFRing~\cite{chen2023efring} utilizes electric field sensing to detect and reconstruct the continuous 1D motion trajectories of the thumb sliding on the index finger. When the thumb approaches the ring, it distorts the electric field generated by the ring, which is detected by the receiving layer of the ring. Similarly, Z-Ring~\cite{waghmare2023zring} detects finger movements by monitoring disturbances in an electric field within its ring-shaped device. As the finger glides over various items and surfaces, the electrical impedance of the hand undergoes alterations. These modifications impact the electric field conducted through the finger, and the Z-Ring can detect the shifts in the reflected signals. It then translates these variations into input signals, thereby reconstructing the finger's trajectory, whether on the hand or traversing a 1D slider crafted from thin copper sheets.

In contrast to electric field methods,  OptiRing~\cite{waghmare2023optiring} employs a miniature endoscopic camera attached to a ring to capture low-resolution images of the thumb and index finger. The images from the camera are processed to monitor changes in the contour area of the fingers within the camera's frame, which helps reconstruct the thumb's relative movement trajectory.

By capitalizing on the fingers' proximity and dexterity and the rings' portability, these methods facilitate high-precision interactions for tasks that demand fine-grained control. Applications range from discreet menu navigation in augmented reality glasses to parameter adjustments in smartwatches and tactile feedback-free control in industrial environments. This technology holds the promise of redefining user interactions with compact devices, emphasizing ergonomic design and seamless integration into daily workflows.

\subsubsection{Two-dimensional Tracking} \label{sec:input_traj_2d}

\textit{Tracking 2D position.}
The IMU is the most straightforward sensor for tracking finger motion data. By wearing a ring with an embedded IMU, MouseRing can accurately detect the 2D trajectory of finger movements.
The Micro Input Devices System (MIDS), as presented by ~\citet{lam2002mids}, serves as a versatile interface input system to substitute traditional computer input devices, including the mouse, pen, and keyboard. Users interact with the system by wearing two MIDS rings, which house commercial MEMS acceleration sensors and a MIDS wristwatch, enabling actions such as clicking and drawing. Furthermore, MIDS incorporates a self-calibration algorithm that addresses motion detection ambiguities encountered by the MEMS sensors.
 ~\citet{younas2020finger} presents a system that uses an IMU-equipped ring to capture finger movements during in-air writing, converting them into 2D trajectories on a digital canvas. It applies a Madgwick filter for sensor fusion to determine the finger’s orientation (roll, pitch, yaw) in the real world, mapping these angles to x and y coordinates on the canvas. Gain values are calculated to translate finger motion into pixel distances, enabling real-time visualization of the writing trajectory. FAirWrite~\cite{younas2022fairwrite} similarly employs an IMU-equipped ring to capture in-air finger movements and reconstruct them into 2D trajectories. However, it places a particular emphasis on the subsequent recognition of these trajectories as alphanumeric characters, thereby facilitating efficient text entry. 
MouseRing~\cite{shen2024mousering} uses a ring with an embedded IMU worn on the finger to track finger movements. It integrates physical priors, such as joint co-planarity and velocity consistency, into ML models to enhance motion perception accuracy. By incorporating touch state detection as a cursor movement switch, it enables precise 2D trajectory recognition and control on various surfaces.

In addition to directly utilizing IMU data for two-dimensional finger motion tracking, continuous two-dimensional position detection can also be achieved through signals generated or altered by the finger's involvement in interactions. One approach involves using the finger as an interaction medium, measuring changes in finger-related signals to track its position. Another method employs proximity sensors to detect the finger's two-dimensional location based on the finger-reflected signals.
Z-Ring~\cite{waghmare2023zring} uses bio-impedance sensing to enable two-dimensional finger position tracking. It leverages active electrical field sensing to detect changes in the hand's electrical impedance caused by finger motions or contact with external surfaces. The ring emits a signal from its electrode pair, which is then received by the same pair after passing through the hand. By analyzing the changes in the received signal's impedance over time and frequency, the system can determine the finger's position in two-dimensional space, enabling precise tracking. SkinTrack~\cite{zhang2016skintrack} is a wearable system that uses the human body as an electrical waveguide to continuously track finger movements on the skin. The system includes a ring that emits a high-frequency AC signal and a wristband equipped with electrodes. By analyzing phase differences between the signals received at different electrodes, SkinTrack can precisely determine the finger's two-dimensional position on the skin, allowing for continuous and dynamic touch tracking.
LightRing~\cite{kienzle2014lightring} is a wearable ring equipped with a single-axis gyroscope and an infrared proximity sensor. It tracks fingertip movement by measuring the rotation angle using the gyroscope and the flexion of the finger by detecting the reflected infrared light intensity as the finger joint bends. With the known values, the system calculates the position of the fingertip in a two-dimensional coordinate system based on the wrist (x, y). 

Other methods, including optimal methods and magnetic field sensing, are also employed for detecting 2D trajectories. Magic Finger~\cite{yang2012magic} is a wearable device that converts any surface into a touchscreen by utilizing an optical mouse sensor to detect finger contact and movement. When worn by the user, the device’s LED illuminates the surface upon contact, and the sensor’s camera captures the reflected light. The system processes these images, identifies key features such as edges and corners, and employs optical flow algorithms to accurately track and calculate the trajectory of the finger’s movement. Abracadabra~\cite{harrison2009abracadabra} leverages magnetic sensing to expand the input area for mobile devices with very small screens, such as smartwatches. By wearing a magnetic ring on the distal phalanx of the index finger, the magnetometer embedded in the smartwatch can measure the magnetic field strength. This measurement indicates the two-dimensional spatial location of the magnet relative to the smartwatch. Their preliminary evaluation of the implemented system achieved a target selection accuracy of 92\%.

\textit{Tracking pointing direction.}
As AR and VR technologies become increasingly prevalent, non-contact pointing interactions with screens have become necessary across numerous applications. Accurate tracking and pointing in the direction of this domain is a crucial subtask. This task is not about tracking the three-dimensional coordinates of the pointing gesture in space, nor is it merely about tracking the fingertip coordinates. Instead, it focuses on tracking the 2D coordinate position on the screen surface projected from the direction the fingertip is pointing. Recognizing the pivotal role of this application, scholars have leveraged diverse sensing technologies to bring it to fruition.

WRIST~\cite{yeo2019wrist} implements pointing direction tracking using a smartwatch with a built-in IMU and a custom smart ring equipped with an IMU sensor. It employs a macro-micro pointing approach where the forearm controls a coarse area cursor, and the wrist or finger controls a precise crosshair. The system maps the wrist and finger movements to cursor positions on the screen, using the forearm motion for coarse positioning and the wrist or finger motion for precise positioning. ~\citet{sibert1997finger} presents a finger-mounted pointing device using a smart ring equipped with an infrared emitter to send signals. Four infrared detectors placed at the corners of a laptop screen receive these signals, and based on Lambert’s cosine law, the system calculates the relative distances from the finger to the screen corners by measuring the signal strength. These distances are then used to compute the Cartesian coordinates of the finger’s pointing position on the screen. ~\citet{das2020real} introduces a novel approach for pointing direction recognition using a ring with an IMU and an RGB-D sensor. The IMU is embedded in a ring on the index finger to obtain its axis vector in the NED coordinate frame. Meanwhile, the RGB-D sensor identifies the hand region via depth image segmentation and locates the fingertip based on distance and angle constraints. The pointing direction is calculated by converting the index finger axis vector into the RGB-D sensor frame and integrating it with the fingertip position. ~\citet{horie2012pointing} proposes a novel interaction method by placing menu options along two intersecting lines on the screen. Users wear rings equipped with IMUs on two fingers. The tilt angles of the accelerometers control the cursor movement along these lines to select the desired menu item.

\subsubsection{Three-dimensional Tracking} \label{sec:input_traj_3d}
The embedded magnet and magnetic sensing technique can be utilized for three-dimensional trajectory tracking. Magnets and magnetometers are typically affixed to the finger and the interacting object (or wrist). As relative motion occurs, the magnet distorts the magnetic field the magnetometer detects. This distortion is leveraged for tracking.

AuraRing~\cite{parizi2019auraring} employs a magnetic ring and a sensor-equipped wristband to track 3D finger movements accurately. The ring generates a fluctuating magnetic field, and the wristband’s sensors detect changes in this field caused by finger motion. Applying physics-based principles and advanced algorithms, AuraRing calculates the ring’s 5-DoF position and orientation with remarkable precision, enabling the system to recognize intricate three-dimensional trajectories with excellent reliability. TRing~\cite{yoon2016tring} also employs magnetic sensing to enable three-dimensional trajectory tracking. Unlike AuraRing, TRing does not have a magnet itself. Instead, magnets are fixed on objects that interact with the hand. The ring’s magnetometer in TRing's IMU detects the magnetic field from these magnets. By combining the magnetic field data with the motion data from the IMU, TRing can achieve real-time three-dimensional tracking of the fingertip's position and orientation, enabling various interactive functions. 
 The MagX~\cite{chen2021magx} system is an advanced wearable hand-tracking solution that employs passive neodymium magnets and a wrist-mounted sensor array to monitor hand movements in three-dimensional space precisely. Users wear rings containing passive neodymium magnets on their index fingers. The sensor array on the wrist comprises multiple magnetometers that continuously detect changes in the magnetic field as the hand moves. The raw data from these magnetometers are processed using sophisticated algorithms, such as the Levenberg-Marquardt method, to accurately calculate the 3D position and orientation of the magnets, thereby reconstructing the hand's trajectory in real time. The uTrack~\cite{chen2013utrack} technique transforms the thumb and fingers into a 3D input system using magnetic field sensing. A user wears a pair of magnetometers attached to the back of the fingers and a permanent magnet placed on the back of the thumb. A continuous input stream is generated as the thumb moves across the fingers, which can be utilized for 3D pointing. The system employs a novel algorithm that directly calculates the magnet's 3D position and tilt angle based on the sensor readings.
 
The main limitation of magnetic field-based 3D trajectory tracking is the tracking distance. TRing can sense magnets up to 1500 mm away but is optimized for accurate monitoring within a 120 × 120 mm area, balancing accuracy with magnetic field strength and environmental noise constraints. AuraRing is also designed for short-range tracking (100–150 mm), which reduces external interference and improves power efficiency. Beyond this range, magnetic field strength decreases, potentially degrading accuracy. Similarly, When the distance between the hand and the sensor array exceeds 300mm, the accuracy of MagX’s tracking significantly diminishes. 

Other methods, including acoustic sensing techniques, visual approaches, and IMU-based methods, can also be used for three-dimensional trajectory tracking. ~\citet{zhang2017soundtrak} introduces SoundTrak, an innovative active acoustic sensing method for continuous three-dimensional finger tracking. SoundTrak employs a miniature speaker on a ring worn on the finger and a set of microphones on a smartwatch. By measuring the phase shifts of the acoustic signals received by the microphones and applying a physics-based model, the system accurately determines the finger's coordinates in three-dimensional space. Like magnetic sensing technologies, this approach has a range limitation. It achieves high-resolution tracking within a 200mm × 160mm × 110mm volume around a smartwatch.
~\citet{das2020real} achieves 3D fingertip localization using RGB-D sensor data. It first segments the hand region from the background using depth thresholding, assuming the hand is the closest object to the camera. Then, it locates the fingertip by searching for the farthest point from the hand's center of mass, ensuring the angle between the line connecting the fingertip to the center of mass and the index finger axis vector is less than 60 degrees. The entire process is optimized for real-time performance using GPU acceleration. ~\citet{Anala2009Simple}introduces a wearable system equipped with IMU sensors, which translates IMU signals into 3D hand position and orientation data through an algorithm analyzing the accelerometer information. While the system described in this study is implemented in a glove, the IMU sensors could also be incorporated into rings, achieving comparable 3D trajectory monitoring capabilities. 3DTouch~\cite{nguyen20153dtouch} utilizes an optical laser sensor(OPS) and a 9-DoF IMU to measure the 3D position relative to a surface. By combining the 2D position relative to the surface provided by the OPS with the 3D orientation provided by the IMU, 3DTouch could derive the relative 3D position. Evaluations showed that the 3D tracking accuracy was 1.1mm.

\subsubsection{Hand Pose Tracking}
Hand pose tracking is a widely explored task with significant applications in fields such as VR/AR, human motion capture, and gesture interaction. However, traditional approaches often require external cameras or complex wearable devices (e.g., multi-sensor data gloves). Achieving continuous hand pose tracking using a single ring can effectively reduce the number of wearable devices, thereby enhancing the comfort of interaction. Nevertheless, due to constraints on the number, size, and power consumption of sensors, this remains highly challenging.
Ring-a-Pose~\cite{yu2024ring} adopts an active acoustic approach. By comparing the transmitted and received signals, a neural network calculates the hand pose, achieving a joint error of 14.1 mm (user-independent) and 10.3 mm (user-dependent).
Z-Pose~\cite{waghmare2023zpose} employs swept-frequency radio frequency (RF) sensing, treating the hand as an RF antenna. By analyzing the emitted and reflected signals, machine learning methods are used to map them to different poses, resulting in a joint error of 7.2 mm (user-dependent).
One Ring to Rule Them All~\cite{zhou2023one} uses IMU data from multiple rings (3–6 rings worn at the bases of fingers or on the wrist) to reconstruct hand poses via deep learning algorithms. Video data is used to generate IMU data for training, ultimately achieving a joint error of 6.57 mm (user-independent, 6 sensors).

\subsection{Text-entry} \label{sec:interaction_input_text}
Text entry is crucial in human-computer interaction, as many human activities involve inputting text. Research on using smart rings for text entry has explored various methods to facilitate this task. Some studies have focused on how to use rings for text or character input. For instance, several studies first constructed a virtual keyboard and then used the ring to detect the typing movements of the hand on the virtual keyboard. These studies recognized the input characters based on the differences in motion data when tapping different keys on the virtual keyboard. Other studies have leveraged handwriting input methods, using the ring to recognize characters the user writes. Additionally, some studies have focused on different populations, detecting sign language with the ring and recognizing and translating it into text to facilitate communication between sign language users and non-sign language users.

\subsubsection{Keyboard.} \label{sec:input_text_keyboard}
Ring-based virtual keyboard serves as an innovative alternative to traditional keyboards, offering portability, ease of use, and potential privacy benefits. It is particularly well-suited for interacting with devices such as VR/AR headsets and smart TVs, providing a seamless and intuitive text input method.
Some smart ring-based keyboard designs are alphabetic, allowing the input of letters, numbers, and symbols. Other studies have implemented a numeric keypad, which can only enter numbers and common symbols.

\textit{Alphabetic Keyboard}
In research on text entry using smart rings for the 26 letters and additional characters, some studies have maintained the basic QWERTY layout. This familiar keyboard layout enhances users' typing efficiency. QwertyRing~\cite{gu2020qwertyring} employs an IMU ring worn on the middle phalanx of the index finger to enable text entry. Users type on any desk-like surface as if using a QWERTY keyboard, with text feedback displayed on a separate screen such as an AR/VR headset or an external monitor. The system detects touch events and predicts desired words based on the IMU ring’s orientation, achieving high accuracy. 
The DRG-Keyboard~\cite{liang2023drg} is designed for typing with the index fingertip. Users wear rings on their thumb and index finger and slide the thumb over an imaginary QWERTY keyboard on the fingertip. This method achieves high accuracy through IMU data and advanced touch detection algorithms, reducing the need for corrections.

Other studies have altered the keyboard layout to facilitate higher accuracy in detection.~\citet{lian2020towards} employs two smart rings worn on the proximal phalanx of one finger of each hand. It features a custom-designed arc-shaped keyboard layout, contrasting with conventional rectangular layouts by enhancing the attitude angle differences between adjacent keys, thereby boosting keystroke recognition accuracy. In addition to the 26 alphabet keys, the character set also includes four function keys: comma, period, space, and delete. The system captures motion data from the ring and achieves a remarkable keystroke recognition accuracy using a k-NN classifier. RingVKB~\cite{li2023ringvkb} implements a virtual keyboard solely by utilizing a single ring worn on the index finger. It features a ring-shaped keyboard with 48 keys, divided into an inner and outer ring, each with 12 sectors. The inner ring has three layers for letters, numbers, and special keys like “Shift” and “Ctrl,” with letters arranged by usage frequency. The system tracks finger movements using an accelerometer and gyroscope to determine keystroke positions. It allows typing in the air, is easy to use, and achieves high accuracy quickly.

Unlike the previous two approaches, ~\citet{meli2017hand} designs a system similar to the T9 keypad, where multiple characters are mapped to a single key to address the issue of accommodating a large set of input characters in a small space.The HAT system comprises multiple rings, each fitted with two conductive bands. Users input characters by tapping different combinations of these bands on the rings and incorporating wrist tilt gestures. Like T9 keyboards, HAT maps multiple letters to a single band, allowing users to select the desired character through tapping frequency or wrist movement.   TypingRing~\cite{nirjon2015typingring} is a wearable ring platform that enables text input into various computing devices by typing on any flat surface. The ring, worn on the middle finger, uses an accelerometer, optical displacement sensor, and proximity sensors to detect hand and finger movements. Users select different zones by moving the ring horizontally and vertically. They then press one of the three fingers (index, middle, or ring finger) to input a specific key. TypingRing supports any keyboard layout and character set, offering versatile input options. Ring GINA~\cite{greenspun2014ring} utilizes a chorded keyboard concept, enabling users to input characters by tapping various combinations of four fingers on their dominant hand, each wearing a ring with an accelerometer. These accelerometers detect taps and generate a four-bit binary signal corresponding to the tapped fingers. Each finger is assigned multiple character options, and users can cycle through and select the desired word using a “Next” action.

\begin{table}[ht]
\centering
\caption{Comparison of input metrics for ring-based virtual alphabetic keyboards}
\begin{subtable}{0.581\textwidth}
\centering
\caption{Efficiency and accuracy of ring-based virtual alphabetic keyboards.}
\resizebox{\linewidth}{!}{%
\begin{tabular}{ccc}
\hline
 & Input Efficiency & Input Accuracy \\
\hline
QwertyRing~\cite{gu2020qwertyring} & WPM=20.59 & UER=1.17\%, CER=3.22\% \\
DRG-Keyboard~\cite{liang2023drg} & WPM = 12.9 & UER = 2.5\%, TER = 4.95\%\\
~\citet{lian2020towards} & - & ACC = 98.53\% \\
RingVKB~\cite{li2023ringvkb} & WPM = 4.56 & ACC = 93\% \\
HAT~\cite{meli2017hand} & CPM = 60 & - \\
TypingRing~\cite{nirjon2015typingring} & KPM ranges from 33 to 50 among users & ACC = 98.67\% \\
Ring GINA~\cite{greenspun2014ring} & WPM ranges from 20 to 30 among users & - \\
\hline
\end{tabular}%
}
\label{subtable:input_keyboard_wpm}
\end{subtable}
\hfill
\begin{subtable}{0.4\textwidth}
\centering
\caption{Explanation of evaluation metrics abbreviation.}
\resizebox{\linewidth}{!}{%
\begin{tabular}{cc}
\hline
Evaluation metrics & description \\
\hline
WPM & words per minute \\
CPM & characters per minute \\
KPM & keystrokes per minute \\
UER & uncorrected error rate \\
CER & corrected error rate \\
TER & total error rate \\
ACC & average keystroke recognition accuracy\\
\hline
\end{tabular}%
}
\label{subtable:input_keyboard_abbr}
\end{subtable}
\label{table:input_keyboard}
\end{table}

\textit{Numeric Keypad.}
As mentioned above, the greatest challenge for alphabetic keyboards with a more extensive set of input characters is to recognize different keys. Therefore, some studies have modified the traditional keyboard layout, extended the keyboard to an external plane, or mapped multiple characters to one key. In comparison, the character set that the numeric keypad can input is smaller and with fewer keys, which is more convenient to integrate into the function of the smart ring: the entire keypad can be implemented in the palm, and touch detection is accurate and easy to implement.
MType~\cite{deng2019mtype} enables text entry by detecting taps on a virtual keyboard projected on the user's hand. The virtual keyboard includes a dial plate with keys {1-9, $\ast$, 0, \#} and a control panel with keys {Return, Up, Delete, Left, Ok, Right, Home, Down, Enter}. The locations of keys can be marked by the texture of the hand. Users wear a magnetic ring and tap the hand, while the smartwatch recognizes keystrokes based on changes in the magnetic field. MType achieves an average accuracy of 93\% with minimal training data, and its performance can be further improved with the runtime adaptation mechanism.

\subsubsection{Handwriting.}

Some studies have explored the use of rings for handwriting text input, employing different sensing technologies to capture and recognize user input.

IMU-based systems leverage IMUs to capture motion data during handwriting. ~\citet{liu2020imu} proposes an on-surface handwriting recognition system on a smart ring. This ring allows users to write on any surface, capturing motion data via an IMU. The data is then transmitted wirelessly to a computer for processing. An LSTM model segments the motion data into individual letters, accurately distinguishing between writing and non-writing states with a 98\% success rate. The segmented data is subsequently recognized as letters or words using either an LSTM or CNN model. The LSTM model demonstrated superior accuracy, achieving a character error rate (CER) of 1.05\% and a word error rate (WER) of 7.28\%, thereby validating the system's potential as a text input interface. FAirWrite~\cite{younas2022fairwrite} enables users to write in mid-air, recognizing their gestures and rendering on a digital canvas. The system reconstructs a real-time 2D trajectory from IMU data. A deep learning model (OS-CNN) then categorizes the reconstructed trajectory into digits (0-9) and letters (A-Z). Additionally, a GUI has been developed to allow users to view the trajectory and manage system functionalities. ~\citet{younas2020finger} has done similar work but focuses more on reconstructing two-dimensional handwriting trajectories.

Before the advent of IMU-based systems, a magnetic field-based wearable handwriting input system was developed. \citet{han2007wearable} introduces a novel approach that utilizes a magnet affixed to a fingertip ring and two sensors mounted on the wrist. The position of the magnet is determined by measuring the magnetic field vector. The system operates in differential mode to effectively eliminate the influence of geomagnetism. Experimental results demonstrate that this method can accurately track the trajectory of handwritten characters with minimal error, thereby providing a viable and innovative solution for wearable text input, particularly in noisy environments or for users with visual impairments.

\subsubsection{Sign Language Recognition.}
Some studies have utilized smart rings to recognize sign language and translate it into text, which can be presented in either written or audio form. This technology enables non-sign language users to understand sign language users, thereby enhancing communication between the two groups. Technically, sign language recognition is a form of gesture recognition. A key challenge is that a comprehensive sign language library for translation and communication requires defining and distinguishing a large number of gestures due to the extensive vocabulary. While some studies have only covered a few frequently used words~\cite{zhang2018fingerping, jing2013magic, kuroki2015remote}, the following two articles have developed a more complete sign language recognition vocabulary database, potentially aiding communication between sign language users and non-sign language users. ~\citet{liu2020finger} presents FinGTrAC, a system that primarily uses motion data from the IMU sensors of a smart ring and a smartwatch to track and translate American Sign Language (ASL) gestures. It combines sparse sensor data with contextual info in a Bayesian framework and applies an HMM for accurate recognition. In testing with 10 participants, the system demonstrated an average recognition accuracy of 94.2\% for 100 commonly used ASL gestures. When the dictionary was enlarged to 200 words, the accuracy slightly dropped to 90\%, demonstrating its high robustness and scalability. The system streams data to an edge device, enabling real-time sentence decoding with a latency of just 0.4 seconds. ~\citet{li2023signring} introduces a wearable solution that employs IMU sensors on smart rings to detect ASL gestures. A key innovation is the creation of virtual IMU data derived from ASL videos, which significantly boosts the size of the training dataset. The system divides continuous sign language input into sentences via a sliding window technique that analyzes the variance in IMU data. Subsequently, a CRNN model with CTC loss identifies the signs within each sentence, managing the segmentation of signs (words) during recognition. It achieves a WER of 6.3\% with mixed IMU and v-IMU training data and can recognize up to 934 unique words and sentences up to 16 glosses in length. Although the study covers a more extensive sign vocabulary compared to FinGTrAC, it did not evaluate real-time performance similarly.

\subsection{Manipulate Ring} \label{sec:interaction_input_mani}
\subsubsection{Discrete Input}


Pointing as a natural indicator of the user's interaction intentions is often used for discrete input such as capturing images by a camera on the ring for visual assistance and interaction with smart devices. 

EyeRing~\cite{nanayakkara2013eyering} presents a ring on the index finger with a camera for both visually impaired people and sighted people. With the camera, it can recognize currency and tags and provide seamless copy-paste interaction with printed media. The experiment shows that EyeRing performs faster than a smartphone application when detecting a single object. The results of the questionnaire show that EyeRing requires less effort and offers hands-free interaction. Other applications built on EyeRing include a brush that captures outlines and colors from the real world to keep children aware of the surrounding world~\cite{hettiarachchi2013fingerdraw}. Similar to EyeRing, Chinese FignerReader~\cite{su2019finger} recognizes Chinese characters on printed material for visually impaired people by a camera and speaks them out. It also contains two vibration motors that guide users' fingers to move along the line. The field test shows that it is useful for reading text outdoors. 
Stearns et al.~\cite{stearns2015design} also used a ring-form-factor device to recognize text on printed material and guide finger movement for visually impaired people. Their prototype raises fewer social acceptability concerns due to the small size of the minimal camera they chose. The image captured by the minimal camera mounted on the middle phalanx is processed using OCR. Feedback was provided by audio, haptic, or both. The result from the user study shows that audio feedback guided reading best.

In addition to reading assistance, other researches focus on helping visually impaired people understand the surrounding world. FingerReader 2.0~\cite{boldu2018fingerreader2} targets at improving shopping experience. With a camera on it, users can get information about products by pointing to them. The on-board deep learning algorithm processes the image first and an external cloud vision API is introduced when the confidence is low. The results from a field study of grocery shopping prove the usefulness of it. However, scalability is an issue because several factors such as objective rotation can influence the result of object recognition and the model was trained on a specific dataset. Imtiaz et al.~\cite{imtiaz2018finger} presented a Smart Eye Assistive device that uses a camera on a ring to recognize the pointed environment and then provides audio assistance. The image transfer time is 6 seconds which reduces its usability.

For smart home interaction, Kim et al. proposed IRIS~\cite{kim2024iris} which integrates camera and IMU for instance-level detection and controlling smart devices while meeting small size, weight, and power (SWaP) requirements. It outperforms voice assistance in user preference and latency. However, the improvement in gesture set scalability and detection accuracy for small objects across far distances could further promote the usability of this ring. Except for the camera, Jing~\cite{jing2013brand} designed a low-memory footprint method that uses a middle device to turn the RF control signal transmitted by a ring into an IR signal to control the home appliances. However, the effectiveness of this method needs future evaluation.


Other works turn the ring into a button to switch between different levels of reality and virtuality. Das et al.~\cite{das2023fingerbutton} designed and compared several ways of transitions between real and virtual environments including button in an index ring (FingerButton), hand gestures, virtual buttons and double taps on the headset. Their conclusion is FingerButton consumes the least trial time and is most preferable. However, its performance in dynamic scenarios is not evaluated and the expansion to switch between more tasks needs more exploration. Considering more states, Das et al.~\cite{das2024exploring} investigated using buttons to transition between four reality-reality states which are real world, augmented reality, augmented virtuality and virtual reality. They discussed the layout of four buttons and compared their solution with a joystick, rotary wheel, and slider. Their study shows that the $2 \times 2$ button is the most effective configuration and built a prototype consisting of two rings based on their findings. Other work that uses the ring as a button includes~\cite{Marti2005giving} which released an egalitarian approach to detect whether to answer a phone call during a meeting using a ring. When a call comes in, every participant receives a vibration notification on the ring and can decide whether to veto by a button.


To achieve quick input when performing other tasks, some works use the ring as a touchpad for navigation and text entry. Bardot et al.~\cite{bardot2021aro} examined two navigation tasks on a smart ring which are zooming and panning in three locations (in-air, on-ring and on-object) while the user is holding an object. The ring prototype consists of a capacitive touch display and an IMU for gesture detection. The results from user studies show that in-air and on-ring locations are more efficient than the on-object location. Boldu et al.~\cite{boldu2018thumb} designed a ring with a touchpad for thumb-to-index gestures for athletic activity. The finding is the most east-to-use gestures are tap, swipe-down, and swipe-left.


Hands-free text entry on a ring can enable seamless and efficient communication in various scenarios. Kim et al.~\cite{kim2018thumbtext} proposed ThumbText, a text entry ring device that supports the input of at least 28 characters by a two-step selection using thumb. The capacitive touch controller on the top of the ring is divided into $2 \times 3$ grids. The user can select a grid by touch and slide. The text entry rates of ThumbText are higher than those of SwipeBoard and H4-Writer. However, the correction, prediction and text edition are not supported. Meli et al.~\cite{meli2017hand} achieves hands-free text entry using a three-rings system. Each ring has two copper bands and a 3-axis accelerometer, which brings multiple combinations of different touching bands and hand orientations. A T9-based text entry method is developed based on this system. The user study shows that the potential text entry rate is at least 60 CPM after training. However, the learning curve indicates the difficulty of learning this system and this method is not capable of entering long text due to its higher error rate and lower entry rate compared to the numpad on a phone.



\subsubsection{Continuous Input}

Several works track the position and rotation of the ring by sensing the changes in the magnetic field. Chatterjee et al.~\cite{chatterjee2006design} proposed a capacitance-sensing-based multi-purpose input device mounted on the fingertip. They achieved both button function and continuous sensing by detecting the change in capacitance. They demonstrate the application of their technology by using their prototype to control a radio controlled car and input Japanese Hiragana characters but the efficiency of their method needs further proof. Ashbrook et al.~\cite{ashbrook2011nenya} developed a ring that supports both selections among 8 targets and “click”. A wrist-worn sensor detects the change of magnetic changes caused by the ring to track the ring’s position and rotation. This ring enables eyes-free operation and also is small and socially acceptable, but the false positive increases when the user is in motion. Cheung et al.~\cite{cheung2019tangible} leveraged the magnetometer and magnets to map the rotation of the ring to angular and linear controls in a mobile game. The results of the user study show that users prefer using the ring and the performance of rotating it for angular control is comparable to touch and surpassing tilt.


Besides, Yau et al.~\cite{yau2020subtle} sought to use a ring to substitute the joysticks for drone control using only one hand. The user study explored the combination of three modalities (force, touch and IMU) and the ring prototype contains a pressure sensor and an IMU for driving the horizontal and rotational drone movements by pitch and roll gestures based on the findings of the user study. The participants found that this ring is easy to use and easy to learn. The quantitative study also shows that their method is faster than the commercial baseline. However, the test on moving the drone on a horizontal plane is not included in this paper.

\section{Interaction - output} \label{sec:interaction_output}

\begin{figure}
    \centering
    \includegraphics[width=0.9\linewidth]{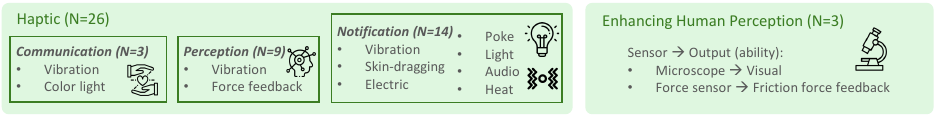}
    \caption{Section outline for interaction output.}
    \label{fig:output_outline}
\end{figure}

The compact form factor of smart rings provides a unique opportunity for discreet and immediate feedback directly to the wearer. The structure of the following section is outlined in Figure~\ref{fig:output_outline}. Section~\ref{sec:output_haptic} details the literature on haptic and other feedback via smart rings for various purposes, including notifications, facilitating human-to-human communication, and providing tactile feedback. Given that human hands are delicate tools capable of performing complex tasks, Section~\ref{sec:output_enhance} explores studies that aim to amplify human abilities and enhance perception.

\subsection{Haptic Feedback} \label{sec:output_haptic}


\begin{table*}[!t]
    \centering
    \caption{Taxonomy of actuators investigated on smart rings based on corresponding feedback provides.}
    \resizebox{1.0\textwidth}{!}{%
    \begin{tabular}{cp{4cm}p{10cm}}
    \hline
    \textbf{Feedback} & \textbf{Actuator} & \textbf{Description}\\
    \hline
    Vibration & Eccentric Rotating Mass (ERM) Actuators, Linear Resonant Actuators (LRA), Voice Coil Actuator (VCA), air bladder & The human skin, particularly the mechanoreceptors in the fingertips, are highly sensitive to vibrations and can distinguish different vibration patterns. Modulating vibration can also replicate the feel of different surfaces in virtual environments. \\
    Skin-drag & Tactor & The DC motor drives the gears, which move the tactor along the ring’s inner surface. The tactor presses against the skin, producing a dragging sensation as it moves.\\
    Poke  & Solenoid valves & When activated, a solenoid extends its shaft, physically poking the wearer's finger. \\
    Pressure \& shear force & Rotating motor & A belt is pulled by two rotating motors. When two motors rotate in opposite directions, a belt presses the finger. When two motors rotate in the same direction, the system generates a shear force.\\
    Light & RGB LED & Colored lighting can evoke users' emotions and thus can be used to promote emotional communication.\\
    Thermal & Resistors & The increasing of temperature can work as a warning and notification. \\
    Sound & Speaker \& buzzer & Auditory cues can be used for message notifications or guiding finger movements for visually impaired people. \\
    Electrotactility & Electrode & By applying small electrical currents, electrotactile systems can directly nerves on skin, creating sensations such as vibration-like effects and texture perception.\\
    \hline
    \end{tabular}
    }
    \label{tab:output_actuator}
\end{table*}

Miner et al.~\cite{miner2001digital} came up with the idea of digital jewelry such as smart rings in 2001. They proposed applications such as notifying the user of messages coming and conveying emotional states to the partner. The subsequent works explored how different actuators can be used to achieve various types of feedback, along with their potential application scenarios. 
Table~\ref{tab:output_actuator} provides a description of all actuators setup used in the literature set. 

\subsubsection{Notification}


Vibration is the most commonly used feedback for notification. Casalino et al.~\citet{casalino2018operator} used a ring to provide vibrotactile notification after the robot detects human intention to improve human-robot collaboration based on the user's head and hand motion. The user study shows that adding haptic feedback significantly reduces the average execution time for non-skilled subjects in assembly tasks. 
Li et al.~\cite{li2024vibrobot} developed a ring with coil actuators for vibration to guide the angular position of finger joints. They designed 6 vibration modes with a recognition accuracy of 96.4\% and used these vibration modes to guide to 10 hand gestures when the user wearing five rings on the hand. The results from the user study show that the average gesture error was reduced from 30 degrees to less than 15 degrees by the guidance of their rings, which has the potential to facilitate hand operation training.
Mayo et al.~\cite{mayo2022development} designed a ring with four linear resonant actuators (LRAs) to generate vibration at the proximal interphalangeal joint. In the experiment, the combinations of two vibration amplitudes, four line thicknesses and two line orientations were tested to identify the perception boundary of a 1D line using vibration feedback. 
Han et al.~\cite{han2017frictio} proposed a passive kinesthetic force feedback for rotational input on a ring. The authors designed six force profiles and the recognition accuracy is 94\%. By mapping different users and states to different force profiles, several applications such as subtle calender and phone call notifications can be enabled. 
Freeman et al.~\cite{freeman2014tactile} experimented with several approaches for giving tactile feedback for above-device gesture interfaces. The ring prototype uses a linear resonant actuator to generate vibration with different frequency components and duration to notify the user of the different stages of gestures. The tactile feedback makes the gesture easier and is preferred by the users.

Some works focus on the notification for accessibility. Stearns et al.~\citet{stearns2016evaluating} compared guiding directional movement of the index finger by audio and haptic feedback on printed material for blind and visually impaired readers. For the haptic feedback, two motors were mounted on the top of the intermediate phalange and the bottom of the proximal phalange on the index finger. The audio feedback is more accurate but haptic feedback is slightly preferred.
Nicolau et al.~\cite{nicolau2013ubibraille} assisted the reading for visually impaired people by mapping the vibration patterns on six rings to the Braille cells according to the metaphor. This method allows the skilled Braille reader to read in private, but the reading rate is low and the comparison with the audio assistance is missing.
Zhang~\cite{zhang2020tapsonic} leveraged an index ring with two linear actuators to assist line chart reading for visually impaired people by combining haptic stimuli and auditory clarification. Users can use swipe and pinch to navigate through different lines and data series. The value difference between data points is presented using bidirectional haptic and the value is read out if the user's finger stays on a data point. The user study shows that the accuracy is up to 85\% and users found this system intuitive to use. The future works include how to expand the method to different types of statistical charts.

Other works explored designing different patterns of skin-dragging for notification. 
Je et al.~\cite{je2017designing} designed and tested skin-dragging patterns on an index ring by changing the speed, direction and length. They found that speed is crucial in determining the implications of motion and leveraging one or two motion parameters is more effective than using all three. Je et al.~\cite{je2017tactoring} stimulated 8 points around the index finger by dragging a small tactor on the skin by a ring (tactoRing) and designed two methods to encode information with 94\% of recognition accuracy. TactoRings can enable applications such as sending notifications from different mobile apps and indicating directions in the navigation app.

Poking could also be used to generate haptic feedback for notification. Je et al.~\cite{je2018pokering} explored using poking to convey information and built a ring prototype (PokeRing) with 8 micro solenoid valves poking at eight different locations around the index finger’s proximal phalanx. They also designed 21 patterns and demonstrated the applications by mapping these patterns to notifications for different apps. However, the prototype needs to be future miniaturized and a comparison with vibration is necessary to show the advantages of poking.

Several forms of new haptics have also been explored.
Stanke et al.~\cite{stanke2020tactilewear} compared electrotactile and vibrotactile feedback using a wristband and a ring and also designed notification patterns for both modalities. The electrotactile brings stronger localized sensations and higher recognition rates for notification, while vibrotactile feedback is more comfortable and less stressful.
Roumen et al.~\cite{roumen2015notiring} investigated the noticeability of five feedback channels—light, vibration, sound, poke, and thermal—across five levels of physical activity (lying down, sitting, standing, walking, and running). The primary conclusion is that vibration is the most reliable and quickest channel for delivering notifications, followed by poke and sound. Light and thermal feedback were found to be less noticeable and more influenced by the level of physical activity.

\subsubsection{Communication}

Through a pair of rings, the changes on vibration and color can be used to promote communication between partners. 
Werner et al.~\cite{werner2008united} embedded a heart monitor and a vibration motor in a ring to deliver intimacy remotely. After measuring the heart rate of the user, the ring transmits it wirelessly to the partner whose ring will simulate the heartbeat by vibration. Participants showed a large interest in this technology in the user study but the interaction and different modes of the ring lack further investigation.
Choi~\cite{choi2014ring} explored using colored lighting and tactile expression to improve remote intimate interactions. When the user presses the button on the smartphone or squeezes the ring with different pressure, the color and vibration intensity on the partner's ring will change.
Similarly, Pradana et al.~\cite{pradana2014emotional} proposed adding vibro-tactile and color-lighting expressions to text messages to promote emotional communication. The findings from the user study prove the effectiveness of invoking emotion using touch and color stimuli. However, the authors didn't experiment with combining the touch and color stimuli.

\subsubsection{Perception}
Haptic feedback is often used to simulate force and tactile for AR/VR interactions.
Because of the small size of the ring, most of works simulate force using two motors on the index finger. When the two motors turn inward at the same time, the user can feel a vertical force. When the two motors rotate in the same direction, the user can feel a shear force. 
Based on this principle, Minamizawa et al.~\cite{minamizawa2008interactive} simulate the mass of the virtual objects in a real box. The experiment on three participants shows that the weight of the objects (10, 50, 100, 200 g) can be recognized by the participants. However, more experiments are needed to explore the efficiency and resolution.
Similar to~\cite{minamizawa2008interactive}, Scheggi et al.~\cite{scheggi2010shape} also included a ring to render the weight of the virtual object by simulating the vertical and shearing force. 
Pacchierotti et al.~\cite{Pacchierotti2016hring} also used the same principle to exert pressure and shear force while not influencing the hand-tracking system such as leap motion. With this ring, performance and perceived effectiveness have improved in the pick-and-place experiment in VR.
Lee et al.~\citet{lee2015enlarging} investigated the just noticeable difference (JND) of visual-proprioceptive conflict and how to reduce it using haptic feedback in VR. The same principle was also used to generate force. 
Different from these, Talhan et al.~\cite{talhan2019tactile} proposed a ring-shaped air bladder (Tactile Ring) that simulates three types of haptic effects: static pressure (up to 6.3 N), high-frequency vibration (up to 2.2 g magnitude and 250 Hz frequency), and an impact (less than 5 ms of latency). 
Tactile Ring can support applications such as surface texture rendering. However, how to render different types of haptic effects at the same time remains unknown.

Textures are commonly simulated by vibration.
Friesen et al.~\cite{friesen2023perceived} simulated textures with three granularities by a ring on the proximal phalanx of the index finger using vibration to avoid decreasing the dexterity of fingertips. The authors compared frequency and amplitude modulation with no modulation and real texture. The main conclusion is that frequency modulation leads to more realistic sensations for coarser textures, while participants were less able to distinguish between modulation types for finer textures.
Similarly, Gaudeni et al.~\cite{gaudeni2019presenting} also used a ring mounted on the proximal phalanx of the index finger to investigate the usability of texture perception of different vibration stimuli. 
Yeom et al.~\citet{yeom2015poster} leveraged vibration feedback to simulate the feeling of writing on a real surface to improve in-air handwriting. The accuracy of character recognition is 92\%. 

Haptic feedback can also be used to convey the user's proximity to objects in VR. Ariza et al.~\cite{ariza2015ringshaped} developed a ring-shaped vibrotactile feedback device for proximity-based cues in VR. They experimented on 5 signal patterns in a 3D virtual object selection task. Their study shows that these vibrotactile feedbacks can improve users' awareness of a virtual objects such as when they penetrate the object or reach the center of the object.

\subsection{Enhancing human perception} \label{sec:output_enhance}
Since the hands are the main part of the body for operating instruments, sensors on the ring mounted on the finger can be used to enhance human perception during operations.
Maeda et al.~\cite{maeda2016hapticaid} leveraged the vibration sensor on the ring and wristband to enhance haptic sensation. The experiment shows that the detection rate of an uneven test tip under a sheet is significantly higher with their system providing haptic feedback. The future applications also include assigning new haptic experiences to the passive objects and communicating haptic experiences.

Enhancing perception by haptic feedback can also be applied to facilitate scientific experiments or even surgery. 
MagniFinger~\cite{obushi2019magnifinger} develops a fingertip camera system that allows users to magnify the image and control the observed area by siding or tilting the index finger. The experiment shows that tilting is better than sliding in both accuracy and movement speed. However, the movable distance is only 0.7 mm, which limits the application of this technology.
Wang et al.~\cite{wang2012portable} present a fingertip-mounted sensor to detect the friction force and torque in real-time for vascular surgery. This paper introduced a force-sensing mechanism to measure the force between the inner shaft and outer socket using 3 off-the-shelf force sensors. However, a new prototype design is needed to reduce the measurement error and device size.

\section{Passive Sensing - In-body Feature} \label{sec:passive_inbody}

\begin{figure}
    \centering
    \includegraphics[width=0.8\linewidth]{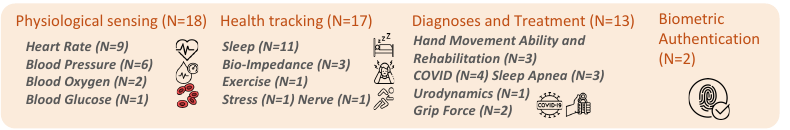}
    \caption{Section outline for passive sensing - in-body feature.}
    \label{fig:inbody-outline}
\end{figure}

The integration of sensing capabilities in smart rings offers a groundbreaking approach to physiological monitoring, enabling continuous tracking of in-body features with minimal intrusion.
The following of this section is structured as Figure~\ref{fig:inbody-outline}. Section~\ref{sec:in-body_cardiovascular} introduces literature focused on the sensing of signals related to the cardiovascular system. Section~\ref{sec:in-body_health} discusses studies aimed at health tracking. Section~\ref{sec:in-body_diagnosis} explores the methods that leverage smart rings for disease diagnose and treatment. explores methods that utilize smart rings for disease diagnosis and treatment. Notably, half of these studies rely on either off-the-shelf ring products or open-source ring platforms, which we summarize in Table~\ref{tab:in-body_commercial}. Additionally, as detailed in Table~\ref{tab:health_gt_description}, we provide a summary of the ground truth methods used in these studies. This summary offers a comprehensive understanding of the quality of the experiments conducted and highlights popular ground truth methods for smart ring sensing. Finally, Section~\ref{sec:in-body_auth} describes biometric-based authentication methods.

\begin{table*}[!t]
    \centering
    \caption{Comparison between commercial smart rings for Health tracking.}
    \resizebox{0.9\textwidth}{!}{%
    \begin{tabular}{|c|c|p{7cm}|c|c|}
    \hline
    \textbf{Device} & \textbf{Sensor type} & \textbf{Function} &\textbf{Price} & \textbf{Paper count} \\
    \hline
    Oura Ring & PPG, IMU, Temperature & 
    Readiness:Daytime Stress, Resilience, Resting Heart Rate, Respiratory Rate, Heart Rate Variability, Body Temperature 
    
    Sleep: Sleep Stages, Body Clock and Chronotype, Blood Oxygen Sensing \((SpO_2)\)

    Activity: Automatic Activity Detection, Daily Movement Graph, Activity Heart Rate

    Heart Health: Cardiovascular Age, Cardio Capacity (VO2 Max), Heart Rate Graph, Live Heart Rate

    Women's Health: Pregnancy Insights, Cycle Insights
    & \$349 & 12\\
    \hline
    RingConn & PPG, IMU, Temperature & Sleep stages and scoring, Blood oxygen levels during daytime and sleep, Real-time heart rate monitoring, Daily activity and calorie consumption, Stress monitoring &\$299 & 2\\
    \hline
    Wellue O2 ring & PPG, IMU & Average \(SpO_2\), Minimum oxygen level, Oxygen drops per hour, Pulse rate, Heart rate, Body Motion & \$179 & 1\\
    \hline
    MoodMetric ring & EDA & stress, emotional level & €529 & 1\\
    \hline
    Senbiosys ring SBF2200 & PPG, Ambient Light Sensor,Proximity Sensor & Blood pressure, Blood oxygen, Heart rate, Heart Rate Variability (HRV), Sleep quality, Sleep quantity, Sleep patterns, Sleep efficiency, Daily activities (steps, calories burned, etc.), Body temperature & Not announced & 1\\
    \hline
    Arcus & IMU & Tracking 3D finger movements, Activity Tracking& Not announced & 1\\
    \hline
    CheckMe O2+ & PPG, IMU & Blood oxygen, Heart rate, Body motion & \$229& 1\\
    \hline
    WristOx2 3150 & PPG & Blood oxygen, Heart rate & \$969.99 & 1\\
    \hline
    OmniRing & PPG, IMU, Temperature &3D finger motion tracking, Heart rate, Blood oxygen, Activity tracking  & open-source & 1\\
    \hline
    \end{tabular}
    }
    \label{tab:in-body_commercial}
\end{table*}

\subsection{Cardiovascular System} \label{sec:in-body_cardiovascular}
The following sections introduce the technical implementation and applications of heart
rate, blood oxygen saturation, blood pressure, blood glucose, and electrocardiography (ECG) sensing on the finger.
\subsubsection{Heart Rate}
\label{sec: Heart Rate}
Heart rate (HR) refers to the frequency of heart contractions and is a key indicator of cardiovascular health. It is typically measured in beats per minute (bpm) by analyzing characteristic waves in PPG or ECG signals. Two common methods for HR measurement include detecting RR intervals (typically measured in ms) for time-domain analysis or applying a Fourier Transform to identify the dominant frequency (usually measured in Hz) in the signal. These techniques provide critical insights into physical fitness and overall health, offering reliable metrics for assessing both normal and pathological conditions.

\textit{Heart Rate - Sensing.} 
Wearable devices in the form of rings have emerged as versatile tools for health monitoring, integrating photoplethysmography (PPG) technology alongside additional sensors. Recent studies highlight the capabilities of ring-based systems for accurate heart rate and inter-beat interval (IBI) detection. For instance, ~\citet{boukhayma2021ring} demonstrated that optimized PPG designs for rings achieve high accuracy even under challenging conditions, with a  97.87\%  beat detection accuracy and IBI estimation errors as low as 8.1 $\pm$ 0.24 ms. ~\citet{haddad2021ear} compared these systems with ear PPG sensors for beat-to-beat interval detection and found that ear sensors performed better (7.52 $\pm$ 0.65 ms). They also evaluated accuracy at reduced sampling frequencies using an ultra-low-power SB200, achieving a 99.27\% detection rate with a mean absolute error of 6.58 ms compared to ECG-derived RR intervals ~\cite{haddad2020beat}. This highlights the sensor’s effective balance between power efficiency and accuracy, making it ideal for long-term and remote HRV monitoring. Such advancements address common issues like peripheral perfusion variability and motion artifacts, enhancing the reliability of rings for continuous monitoring. Moreover, based on the highly accurate detection result above, ~\citet{zhou2020study} developed a ring-type Surgical Pleth Index (SPI) monitoring system which demonstrated 94.75\% agreement with a commercial device (GE Carestation 620 A2) and achieved accurate monitoring with a mean absolute error(MAE) of 2.37 and a bias of -0.13, confirming its reliability under surgical conditions. 

Additionally, ~\citet{spigulis2005optical} compared single- and multi-channel PPG ring devices, finding that single-channel devices consume fewer resources and enable quick acquisition of basic heart rate information, whereas multi-channel devices, capable of accessing different body locations, facilitate synchronized measurements and detailed analysis of pulse wave propagation and vascular resistance transmission. ~\citet{dong2021cloud} incorporated electrodes both inside and outside a ring to measure ECG signals. By positioning the ring at eight different body locations, they achieved 12-lead ECG measurements with remarkable accuracy, demonstrating an RR interval error of only 5 ms compared to the FDA-approved ELI250rx system. Innovative ring-based platforms, such as SensoRing ~\cite{mahmud2018sensoring} and OmniRing ~\cite{zhou2023one}, enhance functionality by integrating sensors like accelerometers, temperature sensors, and IMUs. These advanced systems achieve heart rate measurement accuracies with MAE values of 3 bpm and 8 bpm, respectively, demonstrating their potential for precise and reliable health monitoring.

\textit{Heart Rate - Application.} 
One common application of heart rate measurement is sleep monitoring, facilitated by the Oura Ring and other ring-based devices (see Subsubsection ~\ref{sec: sleep}). Other applications include tracking health-related activities (e.g., exercise ~\cite{koskimaki2019following}) and monitoring body fat (see Subsubsection ~\ref{sec: body fat}).

\subsubsection{Blood Oxygen}
Blood oxygen saturation (SpO2) is a critical indicator of respiratory and circulatory health, reflecting the efficiency of oxygen delivery to the body. Blood oxygen measurement relies on the principle that hemoglobin absorbs light differently depending on its oxygenation state. By emitting light at specific wavelengths, such as red light (e.g., 660 nm) and infrared light (e.g., 880 nm), and measuring the relative absorption, the blood oxygen saturation can be calculated. This non-invasive method, often implemented in photoplethysmography (PPG) devices, provides a convenient and reliable way to monitor SpO2 levels. ~\citet{magno2019self} designed a self-charging ring to measure blood oxygen but only conducted a single-subject and breath-holding study to validate that it has a similar trend with Nellcor N-395 Oximeter. Moreover, ~\citet{santos2022use} assessed wearable ring-like pulse oximeters, including CheckMe O2+, AP-20, and WristOx2 3150, especially the performance during motion and highlighted the significant impact of motion artifacts on SpO2 measurement. In regard to this dilemma, ~\citet{magno2019self} incorporated an accelerometer into their design, aiming to mitigate motion artifacts in future iterations. The current limitations during motion underscore the need for further refinement in wearable ring-like SpO2 technology, requiring additional sensors and more advanced signal processing algorithms.

\subsubsection{Blood Pressure}
Accurate blood pressure (BP) monitoring is essential for effective health management, and recent advancements have introduced wearable devices such as rings and wristbands as non-invasive solutions. The development of ring-based blood pressure (BP) monitoring systems has significantly progressed over the years, demonstrating improved accuracy and practicality. Early work by ~\citet{song2009new} introduced a novel cuff unit designed for instantaneous BP measurement at the finger artery. This method utilized localized pressurization to capture real-time BP readings, demonstrating the feasibility of finger-based monitoring, albeit with limitations in subjective diversity. A decade later, ~\citet{liu2019feasibility} investigated fingertip oscillometric BP measurement, identifying the distal phalanx as an ideal site. Their study achieved promising accuracy, with systolic BP errors averaging \( -6.7 \pm 12.9 \) mmHg and diastolic BP errors averaging \( 1.7 \pm 8.7 \) mmHg, further solidifying the potential of non-invasive finger-based BP monitoring.

More recent advancements have focused on leveraging photoplethysmography (PPG) and bio-impedance technologies to enhance performance and usability without requiring cuff pressure. ~\citet{liu2018multi} proposed a depth-resolved multi-wavelength PPG (MWPPG) method with arteriolar pulse transit time (PTT) for cuffless and continuous blood pressure monitoring, which utilizes four different LED wavelengths to reconstruct arterial pulse PPG from any three MWPPG signals and defines the shortest wavelength PPG as the capillary pulse. This approach achieved high accuracy, with the mean absolute difference (MAD) of 2.85 mmHg for systolic BP and 1.75 mmHg for diastolic BP, representing a significant advancement that enables precise BP and systemic vascular resistance (SVR) measurements with a single, compact wearable sensor. Moreover, ~\citet{panula2020instrument} developed the FANTOM system, a novel cuffless tonometric device for fingertip blood pressure monitoring, which innovatively combines oscillometric and tonometric techniques to measure both peripheral and central blood pressure (CBP) directly from the fingertip. It achieves high accuracy with mean errors of \(-0.9 \pm 7.3 \) mmHg for systolic BP, \(-3.3 \pm 6.6 \) mmHg for diastolic BP, and \(-5.8 \pm 3.2 \) mmHg for CBP compared to gold-standard devices. Additionally, it estimates central augmentation index (cAIx) with an error of \(1.4 \pm 6.2\% \), providing a comprehensive assessment of arterial stiffness. This compact and user-friendly integration design underscores its potential for portable cardiovascular health monitoring. 

From another perspective, ~\citet{paliakaite2021blood} compared finger-based and wrist-based PPG signals for blood pressure (BP) estimation, demonstrating that finger-based models are more effective in capturing BP trends during cold pressor tests. For systolic BP, the mean ± standard deviation of differences between reference and estimated values was \( 0.47 \pm 10.44 \) mmHg for finger PPG signals and \( 1.05 \pm 12.86 \) mmHg for wrist PPG signals. The MAD was 7.78 mmHg for finger PPG signals, compared to 9.69 mmHg for wrist PPG signals, underscoring the superior accuracy of finger-based models. However, changes in systolic BP of at least 10 mmHg were detected with F1 scores of 0.81 for finger PPG and 0.76 for wrist PPG. Complementing this, ~\citet{osman2022blood} developed a bio-impedance ring sensor designed to detect arterial blood flow in the finger with remarkable precision. Their device achieved average MAE values of \( 4.38 \) mmHg for systolic BP and \( 3.63 \) mmHg for diastolic BP, with \( 90\%\) and \( 93\%\) of predictions, respectively, falling within an error margin of \( \pm 10 \) mmHg. These results align with the accuracy standards set by the Association for the Advancement of Medical Instrumentation (AAMI). Such advancements mark a significant leap from foundational technologies to highly accurate, wearable solutions, paving the way for more user-friendly and portable BP monitoring systems.

\subsubsection{Blood Glucose}
Recent advancements have explored non-invasive blood glucose monitoring using wearable devices like rings. ~\citet{cruz2019application} investigated the feasibility of using reflectance-mode PPG sensors integrated into a ring to measure blood glucose levels. Their study demonstrated a significant correlation between PPG peak voltages and blood glucose levels, achieving a positive linear relationship (\( R^2 = 70.3\%\)) for non-diabetic individuals and a negative linear relationship (\( R^2 = 91.92\%\)) for diabetic individuals. While promising, this research highlights the need for further refinement and validation to transition from exploration to practical application in everyday health monitoring.


\subsection{Health Tracking} \label{sec:in-body_health}
Ring devices have become essential tools for health tracking, providing continuous and non-invasive monitoring of key physiological metrics. The following sections discuss their applications in sleep monitoring, bioelectrical impedance analysis, exercise tracking, stress detection, and nerve activity assessment.

\subsubsection{Sleep}
\label{sec: sleep}
During sleep, individuals are exposed to fewer external disturbances and maintain a prolonged resting state, enabling precise detection of dynamic physiological changes such as heart rate (HR), heart rate variability (HRV), respiratory rate (RR) and skin temperature. This facilitates a more accurate and effective assessment of overall health. Meanwhile, given that fingers' rich vascular network ensures stable PPG signals and is highly sensitive to temperature, ring devices are ideal for enhancing sleep measurement accuracy with unobtrusive design and long wearability.

Currently, most researchers prefer to use commercial ring devices, especially the Oura ring, which, as mentioned in \ref{sec: Heart Rate}, can reliably measure HR and HRV during sleep. ~\citet{kinnunen20180312} compared these two metrics with ECG and reported a high correlation for HR (\(r^2\) = 0.998, bias = -0.53 bpm) and HRV (\(r^2\) = 0.967, bias = -0.47). As for overall performance, Oura ring has been compared to polysomnography (PSG) for sleep tracking, achieving 96\% sensitivity in sleep detection. However, it showed limitations in wake specificity (48\%) and detailed stage accuracy, particularly in deep sleep (N3)~\cite{de2019sleep}. Meanwhile, ~\citet{rajput2023assessment} evaluated the Oura ring in free-living conditions and found a high correlation with sleep diaries(Spearman’s r = 0.75), demonstrating its reliability for monitoring total sleep duration in real-world settings. Moreover, when combined with other devices and supervised learning algorithms, Oura ring enhances sleep tracking performance. For instance, ~\citet{malakhatka2021monitoring} utilized ring data along with smart building sensors to predict sleep quality by analyzing both physiological metrics and environmental factors like temperature and humidity. While ~\citet{kazemi2023can} used an Extreme Gradient Boosting model with ring's sensor data to predict sleep quality attributes and achieved high accuracy (90.58\% for sleep duration, 95.38\% for efficiency, 91.45\% for deep sleep).

In addition to assessing physiological metrics, Oura Ring integrates proprietary algorithms and provides an app-based feedback system to deliver an accurate assessment of sleep quality and quantity. By analyzing data such as total sleep time, sleep stages and disturbances, it generates numerical evaluations and feedback, making it a comprehensive software-hardware tool for sleep assessment. ~\citet{koskimaki2018we} used Oura Ring app to examine the relationship between sleep duration, consistency, and timing on sleep quality and found that consistent sleep schedules correlated with longer sleep duration and higher efficiency. ~\citet{peltonen2022impact} investigated the impact of smartphone usage on circadian rhythms and sleep quality using Oura Ring app and employed machine learning models to predict sleep parameters, achieving prediction accuracies ranging from 31\% to 67\%. Moreover, ~\citet{kuosmanen2022does} examined how sleep tracking with Oura Ring app influenced daily habits and perceptions of data reliability. Their findings highlighted both positive behavioral changes, such as increased sleep awareness, and challenges, including stress associated with constant tracking.

While the Oura ring dominates sleep tracking research, other commercially available rings, including RingConn and Wellue O2, are also used for specific sleep-related applications. Recent advancements indicate that these other commercial ring devices are primarily utilized to detect sleep apnea (see subsubsection \ref{sec: sleep apnea}) and have demonstrated strong performance in detection\cite{wongtaweesup2023using}\cite{guo2024osahs}\cite{wu2024transformer}. These developments underscore the versatility of ring devices as comprehensive tools for general sleep monitoring, targeted health tracking and interventions.

\subsubsection{Bioelectrical Impedance} \label{sec: body fat}
Bioelectrical impedance is a technique used to evaluate body composition by measuring impedance, which correlates with physiological parameters such as fat mass, hydration and cellular health. In recent years, its integration into wearable ring devices has emerged as a potential solution for monitoring body composition and related health metrics. For instance, ~\citet{usman2018ring} developed a four-electrode ring-based bio-impedance analyzer for body fat estimation, demonstrating a strong correlation with commercial analyzers (r = 0.9), highlighting its potential as a portable and reliable alternative for fat mass assessment. Expanding on this, ~\citet{usman2019analyzing} compared dry copper electrodes integrated into the ring with traditional wet Ag/AgCl electrodes and, after mitigating noise levels through signal processing, achieved high correlation coefficients (resistance r = 0.96, reactance r = 0.93).

Beyond body composition, ~\citet{osman2022blood} extended the application of BIA by developing a single-channel bio-impedance ring sensor for blood pressure estimation and achieved errors of 4.38 mmHg (systolic) and 3.63 mmHg (diastolic) with 90\% of measurements falling within ±10 mmHg. These advancements underscore the versatility and scalability of ring devices in bioelectrical impedance while highlighting the potential for future improvements in accuracy and noise reduction.

\subsubsection{Exercise and Alcohol}
As described in ~\ref{sec: Heart Rate}, physiological metrics such as resting heart rate (RHR) and heart rate variability (HRV) provide valuable insights into the balance between physical strain and recovery, offering actionable feedback for optimizing exercise routines or managing the effects of alcohol intake, both of which significantly impact the body. Wearable ring devices serve as powerful tools for non-invasive, long-term monitoring, capturing nocturnal physiological data with high resolution to assess recovery status. Notably, ~\citet{koskimaki2019following} used Oura Ring to investigate how both exercise and alcohol consumption influence RHR and HRV. Their findings showed that less-trained participants experienced significant HRV reductions (up to 25 ms) after exercise, while trained individuals exhibited minimal changes, indicating greater physiological resilience. The study also observed that alcohol intake consistently led to increased RHR and decreased HRV, underscoring the utility of ring-based nocturnal measurements in distinguishing the acute effects of these behaviors.

\begin{table*}[htbp]
    \centering
    \caption{Application/Ground Truth Method Types and Descriptions for Health}
    \resizebox{0.9\textwidth}{!}{
    \begin{tabular}{>{\columncolor{gray!20}}l p{4cm} p{11cm} c}
        \multicolumn{2}{l}{\textbf{Application/Ground Truth Method}} & \textbf{Description} & \textbf{Paper Count} \\
        \hline
        & Oura Ring & PPG, accelerometer and temperature sensors track heart rate, movement and body temperature to assess sleep quality and physiological trends. & 2 \\
        \cdashline{2-4}[1pt/1pt]
        & PSG & Record brain activity, eye movements, muscle activity, heart rate and respiration to assess sleep stages and detect abnormalities. & 4 \\ 
        \cdashline{2-4}[1pt/1pt]
        \multirow{-3}*{\raisebox{3mm}{\rotatebox{90}{\makecell{Sleep}}}} & Self-Assessment \& Surveys & User feedback on their own behavior, physiological sensations, and psychological states. &3 \\ 
        \hline
        & Accelerometer & Measure inertial acceleration on one or more axes to track body motion, posture changes, and activity levels. & 3 \\ 
        \cdashline{2-4}[1pt/1pt]
        \multirow{-2}*{\rotatebox{90}{\makecell{Motion}}} & Infrared Cameras & Detect infrared radiation to capture body movement, temperature changes and physiological signals in no-contact conditions. &2 \\ 
        \hline
        & ECG & Measure electrical activity of the heart to monitor heart rate, rhythm and detect cardiovascular abnormalities. & 2 \\ 
        \cdashline{2-4}[1pt/1pt]
        \multirow{-2}*{\rotatebox{90}{\makecell{Heart\\Rate}}} & PPG & Use light absorption changes in blood vessels to measure heart rate, blood oxygen levels and circulatory dynamics. & 8 \\
        \hline
        & Finger Cuff (PPG \& Pressure Sensor) & Combine PPG and pressure sensing to measure continuous blood pressure through arterial volume changes. & 2 \\ 
        \cdashline{2-4}[1pt/1pt]
        & Finger Cuff (Pressure Sensor only) & Apply controlled pressure on the finger artery to measure continuous blood pressure and vascular properties. &2 \\
        \cdashline{2-4}[1pt/1pt]
        & Upper Arm Blood Pressure Monitor (Oscillometric Method) & Measure blood pressure by inflating a cuff and analyzing pressure oscillations in the artery during controlled deflation. & 3 \\ 
        \cdashline{2-4}[1pt/1pt]
        & Pulse Wave Analysis & Analyze pressure waveforms in arteries to assess vascular health, arterial stiffness, and central blood pressure. &1 \\
        \cdashline{2-4}[1pt/1pt]
        \multirow{-5}*{\rotatebox{90}{\makecell{Blood pressure}}} & Arterial Blood Gas Analysis & Use arterial catheterization to measure blood pressure along with other metrics, providing insights into cardiovascular and respiratory function. &1 \\
        \hline
        & Blood Glucose Meter & Measure daily blood glucose levels using a small blood sample, typically via an enzymatic electrochemical sensor. & 1 \\ 
        \cdashline{2-4}[1pt/1pt]
        \multirow{-2}*{\rotatebox{90}{\makecell{Blood\\Glucose}}} & Fasting Blood Glucose Test & Measure blood glucose levels after an overnight fast to assess baseline glucose regulation and detect diabetes risk. & 1 \\
        \hline
        & Pulse Oximeter (Transmissive PPG) & Use light transmission through tissue to measure blood oxygen saturation by analyzing changes in light absorption at specific wavelengths. & 1 \\ 
        \cdashline{2-4}[1pt/1pt]
        \multirow{-2}*{\rotatebox{90}{\makecell{Blood\\Oxygen}}} & Blood Gas \& Electrolyte Analyzer & Measure blood gases, pH, electrolytes and metabolites using ion-selective electrodes and optical sensors to assess respiratory function. & 1 \\
        \hline
        & Force Torque Sensor &  Measure applied force and torque on multiple axes using strain gauges to analyze grip strength and mechanical interactions. & 1 \\
        \cdashline{2-4}[1pt/1pt]
        & Piezoelectric Sensor & Detect mechanical stress or pressure by converting it into an electrical signal, enabling measurements of force and vibration. & 2 \\ 
        \cdashline{2-4}[1pt/1pt]
        \multirow{-3}*{\rotatebox{90}{\makecell{Force}}} & Handgrip Dynamometer & Measure grip strength by detecting the force applied to a calibrated spring, assessing muscular strength and fatigue. & 1 \\ 
        \hline
        & Ag/AgCl Electrodes & Provide stable, low-noise signal acquisition with excellent biocompatibility, commonly used for ECG, EEG and bioimpedance measurements. & 1 \\ 
        \cdashline{2-4}[1pt/1pt]
        \multirow{-2}*{\rotatebox{90}{\makecell{Bio-Im\\pedance}}} & Stainless Steel Electrodes & Provide durability and reusability but may have higher impedance and lower signal stability, used in bioimpedance and EDA sensing. & 1\\
        \hline
        
        \raisebox{-4mm}{\rotatebox{90}{EDA}} & Ag/AgCl Electrodes & Provide stable, low-noise signal acquisition with excellent biocompatibility, commonly used for ECG, EEG and bioimpedance measurements. & 1\\
        \hline
        \raisebox{-4mm}{\rotatebox{90}{\makecell{Urine}}} & Conventional Uroflowmeter & Measure urine flow rate and volume over time using a weight-based or rotating disk sensor to assess urinary tract function and detect abnormalities. & 1\\
        \hline
        \raisebox{-4mm}{\rotatebox{90}{\makecell{Stress}}} & Self-Assessment \& Surveys & User feedback on their own behavior, physiological sensations, and psychological states. & 1\\
        \hline
        & Self-Assessment \& Surveys & User feedback on their own behavior, physiological sensations, and psychological states. & 3 \\ 
        \cdashline{2-4}[1pt/1pt]
        \multirow{-2}*{\rotatebox{90}{\makecell{Covid}}} & Clinical Evaluation \& Medical Reports & Assess physiological and psychological conditions through standardized medical tests, professional observations, and diagnostic reports by healthcare providers. &3 \\
        \hline
        & Self-Assessment \& Surveys & User feedback on their own behavior, physiological sensations, and psychological states. & 1 \\ 
        \cdashline{2-4}[1pt/1pt]
        \multirow{-2}*{\rotatebox{90}{\makecell{Rehab-\\ilitation}}} & Clinical Evaluation \& Medical Reports & Assess physiological and psychological conditions through standardized medical tests, professional observations, and diagnostic reports by healthcare providers. &1 \\
        \hline
    \end{tabular}
    }
    \label{tab:health_gt_description}
\end{table*}

\subsubsection{Stress}
Stress tracking is a critical aspect of health monitoring and wearable rings equipped with skin conductance (SC), electrodermal activity (EDA), and galvanic skin response (GSR) sensors provide effective solutions. The finger’s sensitivity to autonomic nervous system changes enables ring devices to precisely detect physiological stress responses. ~\citet{halkola2019towards} used the MoodMetric ring to monitor stress across six gameplay phases, finding significant correlations between SC, EDA, and GSR data with stress variations, such as increased stress during high-difficulty phases and decreased stress in cool-off periods. In contrast, self-reported stress showed no significant correlation, further validating the reliability of ring devices for objective stress monitoring in dynamic environments.

\subsubsection{Nerve}
A firm grasp of the autonomic nervous system (ANS) is crucial for understanding how the body regulates essential functions such as stress responses and cardiovascular activity. Continuous ANS monitoring offers valuable insights into physiological and emotional well-being, aiding in the early detection of stress and fatigue. The finger, being highly vascularized and sensitive to autonomic changes, serves as an ideal site for measuring electrodermal activity (EDA), heart rate (HR), skin temperature and motion. ~\citet{mahmud2018integrated} developed SensoRing, a compact, power-efficient and unobtrusive solution for ANS tracking, which demonstrated HR deviations under 3 beats per minute(bpm) and strong EDA correlations with the ground truth (0.94 < r < 0.99). The system also achieved over 90\% accuracy in mean RR intervals (775.4 ms) and standard deviation (109.47 ms) with power consumption ranging from 2.9–8.3 mA. This multi-sensor approach enables comprehensive monitoring of ANS activity and stress responses. By combining precision with ease of use, wearable ring devices offer a reliable and non-invasive method for continuous ANS tracking, positioning themselves as promising tools for personalized health management.

\subsection{Diagnosis and Treatment} \label{sec:in-body_diagnosis}
Ring devices enable diverse applications in diagnosis and treatment by leveraging their finger-mounted design and integrated sensors. The following sections explore their roles in grip force monitoring, hand movement ability, rehabilitation, urodynamics measurement, sleep apnea detection and COVID-19 management.

\subsubsection{Grip Force}
It's well established that grip force is a critical health metric widely used to assess various aspects of health status, which is closely linked to various clinical and health outcomes, including frailty in older adults, cardiovascular health and rehabilitation progress. Meanwhile, recent innovations in wearable ring devices offer a portable, non-invasive alternative for continuous grip force monitoring through various sensors integrated into rings. ~\citet{urban2013computing} utilized a ring-mounted camera to detect nail coloration and deformation, predicting torque in the XYZ directions through optical appearance analysis to estimate the grip force. From another aspect, ~\citet{yin2023hippo} introduced HIPPO, which uses embedded light sensors to detect grip-induced deformations on object surfaces,  and achieved 86\% accuracy in estimation and classification. These advancements highlight the potential of ring devices for continuous grip force monitoring and enabling seamless integration into natural behaviors.

\subsubsection{Hand Movement Ability and Rehabilitation}
Given the high flexibility and frequent use of hand joints in daily activities, finger-worn ring devices equipped with IMU sensors are uniquely positioned to monitor both subtle hand movements and broader upper-limb dynamics. This capability further extends their application to medical rehabilitation assistance for conditions involving finger mobility impairments.
Recent advances indicate that researchers have explored the use of ring devices to monitor hand movement ability and provide targeted support for rehabilitation training. ~\citet{liu2017novel} designed a finger-worn sensor with a 9-axis IMU to estimate hand activity levels by quantifying movement patterns, achieving high reliability (ICC = 0.84). Their system effectively detected differences in activity levels before and after rehabilitation. Furthermore, shifting the focus to rehabilitation, ~\citet{liu2018use} managed to monitor bilateral hand and arm movements during stroke rehabilitation, measuring key metrics like motion intensity, usage duration and inter-limb ratios, which showed significant correlations with clinical benchmarks \((Pearson\ r = 0.78, p < 5.6 × 10^{-204})\). Specifically, ~\citet{kim2019towards} developed an mHealth system using finger-worn IMU sensors to monitor motion intensity, usage duration, and inter-limb ratios, demonstrating strong correlations with clinical benchmarks(r = 0.83, p < 0.01) and offering a more comprehensive approach to tracking rehabilitation progress. These studies pave the way for further research to enhance rehabilitation strategies with hand movement detection and recovery outcomes.

\subsubsection{Urodynamics Measurement}
Urodynamics provides valuable insights into bladder function during storage and voiding. While traditional methods often rely on immobile, toilet-mounted devices, which can be cumbersome and limited to clinical environments, wearable devices such as rings present a promising alternative for collecting urodynamic data. These devices offer convenient and portable monitoring solutions, enabling the assessment of flow rate, voiding volume and voiding patterns. ~\citet{matsumoto2012wearable} developed a fingertip-mounted device which utilizes a 40 kHz airborne ultrasound Doppler system. Leveraging the strong echogenicity of water droplets to ultrasound waves, the device measures Doppler-shifted signals caused by running urea (water) drop in air and captures instantaneous flow rates and voiding patterns. Despite challenges in calibrating Doppler signals and ensuring accuracy across varying conditions and patient profiles, wearable devices like fingertip-mounted Doppler sensors demonstrate the potential to revolutionize urodynamic measurements. By providing a patient-friendly, mobile and natural monitoring option, such ring-like devices pave the way for more accessible urodynamics measurement.

\subsubsection{Sleep Apnea}
\label{sec: sleep apnea}
Generally, sleep apnea is defined as a cessation of breathing for 10 seconds or more during sleep ~\cite{ferri2009ferri}. Since sleep apnea is classified by an oxygen saturation drop of at least 3\% within 10 seconds and will alter PPG signal absorption, ~\citet{wongtaweesup2023using} tested the Wellue O2 ring to classify Obstructive Sleep Apnea (OSA) severity using whole-night blood oxygen saturation data. The ring achieved an approximate 65\% detection rate compared to gold-standard polysomnography (PSG) and significantly outperformed smartwatches, though its accuracy in estimating OSA severity based on apnea frequency was limited. Building on this, ~\citet{guo2024osahs} evaluated the RingConn smart ring, which integrates 3-channel PPG, a 3-axis accelerometer and a temperature sensor. By utilizing multi-modal data from PPG and SpO2 signals, it achieved a strong correlation with PSG-derived AHI (r = 0.93) and AUROC values of 0.80 \((AHI \geq 5)\), 0.83 \((AHI \geq 15)\) and 0.74 \((AHI \geq 30)\), highlighting its enhanced accuracy for OSAHS detection. While
~\citet{wu2024transformer} used the same tools but developed a transformer-based model, rather than hardware improvement, to analyze PPG and SpO2 signals, achieving a strong correlation with PSG-derived AHI \((\rho = 0.96)\) and AUROC of 0.83 \((AHI \geq 5)\). These studies collectively highlight the potential of ring devices as practical, non-invasive and patient-friendly alternatives for early detection and screening in sleep apnea.

\subsubsection{COVID}
Retrospective data indicate that COVID-infected patients often exhibit autonomic nervous system disturbances, including elevated Resting Heart Rate (RHR) and reduced Heart Rate Variability (HRV), along with dyspnea or abnormal respiratory rates, with fever being also a typical early symptom. These physiological changes make ring devices efficient for health monitoring in COVID. By integrating PPG and temperature sensors, ring devices deliver high-quality data, leveraging the finger’s rich vascularity for stable PPG signals and heightened sensitivity to core body temperature. Based on the physiological data collected from Oura ring, including HR, HRV, RR and skin temperature, ~\citet{smarr2020feasibility} explored its feasibility in fever monitoring, found that skin temperature data effectively reflected fever onset with an average increase of \(+0.63^o C\) (p=0.024) and 93\% of cases had a fever-like abnormality within seven days before symptom onset. While ~\citet{joseph2022integrated} utilized these data to develop an Ensemble Bayesian Deep Learning Classifier (EBDLC) for the early detection of asymptomatic COVID cases. Their model classified risk levels into low, medium, and high, achieving an accuracy of 97\%, sensitivity of 97.4\%, specificity of 97.2\% and precision of 98\%. Meanwhile, ~\citet{poongodi2022diagnosis} focused on RNN and CNN-based models to rapidly predict COVID infections within 24 hours, achieving an accuracy of 96.8\% in distinguishing between early and severe stages based on clinical thresholds.

Beyond identifying more effective metrics for the early recognition of asymptomatic or mild infections, preventing and controlling pandemics like COVID requires systems capable of long-term health monitoring. ~\citet{purawat2021tempredict} developed a comprehensive data-centric platform named TemPredict to integrate data collection, preprocessing, feature extraction and advanced neural models to detect pre-infection anomalies, assess COVID risks and analyze symptom dependencies. In terms of expansion, the potential applications of ring devices extend beyond COVID, positioning themselves as valuable tools for public health and infectious disease management in the future

\subsection{Biometric Authentication} \label{sec:in-body_auth}
Smart rings have emerged as a promising platform for user authentication by exploiting the unique biological and anatomical features of the wearer. Recent advancements in this domain have leveraged various sensing modalities to enhance security and user convenience. These proposed methods offer robust security against malicious attacks, as the unique composition and biological features of fingers are difficult to replicate or preserve if severed. This makes smart rings an attractive option for biometric authentication in various applications.

Relying on multi-frequency impedance sensing with hand acting as antenna, Z-Ring~\cite{waghmare2023zring} authenticates users by leveraging their unique hand composition and anatomical structure, which can be characterized as an impedance signature. A challenge with using electric field sensing is that touching external surfaces can alter the sensed impedance, potentially disrupting always-available authentication methods. To address this, Z-Ring incorporated six commonly used objects as potential distractions in their experiment. Employing a Random Forest Classifier, they achieved 99\% accuracy in user identification and 98\% accuracy in user authentication. 
~\citet{iwakiri2023user} utilize active acoustic sensing with a microphone and speaker setup positioned on opposite sides of a ring to implement an automatic and continuous user authentication method. Similar to the electric field sensing approach, the unique shape and composition of a user's hand act as a biometric passcode. This method operates by emitting a sine wave sweep signal and then calculating the Dynamic Time Warping (DTW) distance between the received frequency power spectrum and the registered one. To assess the robustness of this authentication method, the study conducted experiments in both relaxed and gripping hand states, achieving approximately 97\% accuracy.

Authentication methods other than biometric-based approaches are summarized in Section~\ref{sec:outbody_authentication}.


\section{Passive Sensing - Out-body Feature} \label{sec:passive_outbody}

\begin{figure}
    \centering
    \includegraphics[width=0.85\linewidth]{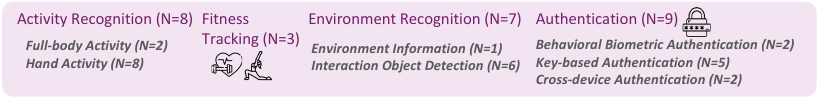}
    \caption{Section outline for passive sensing - out-body feature.}
    \label{fig:outbody-outline}
\end{figure}

Smart rings equipped with sensing technologies extend their utility beyond physiological monitoring to include the detection and analysis of out-body features, such as monitoring physical activity and understanding environment information. The following of this section is structured as Figure~\ref{fig:outbody-outline}. Section~\ref{sec:outbody_activity} details literature that focuses on monitoring users' daily activities. Section~\ref{sec:outbody_fitness} introduces studies that use smart rings as a platform for fitness tracking. Section~\ref{sec:outbody_environment} explores studies aimed at understanding environmental information, including environmental events and objects with which the user interacts. Finally, Section~\ref{sec:outbody_authentication} introduces authentication principles that rely on external features, contrasting with the biometric-based methods discussed in Section~\ref{sec:in-body_auth}.


\subsection{Activity Recognition} \label{sec:outbody_activity}
Smart rings serve as an ideal medium for activity recognition. On one hand, rings are easy to wear and do not interfere with daily life, allowing continuous daily monitoring. On the other hand, the hand, where the ring is worn, is one of the most agile parts of the human body and is involved in a wide range of activities. A smart ring can effectively detect both full-body activities and fine-grained hand activities. The previous one involves activities that engage large muscle groups across the body, such as running, walking, stand-to-sit, and sit-to-stand, while the latter one focuses on the intricate movements of the hands and fingers., including activities such as typing, writing, manipulating utensils while eating, and interacting with a steering wheel while driving.  



\subsubsection{Full-Body Activity Recognition}
Full-body activity recognition using a smart ring is typically achieved through IMUs. As a result, only dynamic movements can be detected, while static postures such as sitting, lying down, or standing still cannot be distinguished using this method. 
~\citet{liu2017novel} designs a wearable system that combines a wristband and a ring. By synchronously detecting hand movements using IMUs on both the wrist and the finger, this system can recognize three gross motor activities: walking, sit-to-stand, and stand-to-sit. In contrast, ~\citet{zhou2015healthcare} uses a single ring with an IMU to monitor activities, identifying running, walking, and jogging. They further categorized jogging into morning and evening jogging based on the time of the activity.
In addition to gross motor activity recognition, both studies also involved different fine motor activity recognition types. Although the granularity of these two types of recognition may seem inconsistent, the precision and sensitivity of the ring IMU allow both tasks to be performed without the need for task separation.

\subsubsection{Fine-Grained Hand Activity Recognition}
Fine motor activity recognition has been explored in various contexts, with some studies focusing on non-object-interacting activities while others concentrate on object-interacting activities.
In non-object-interacting activities, ~\citet{xu2022washring} proposes a medium-based handwashing detection method utilizing the IMU of a waterproof Oura Ring. This approach can recognize nine handwashing steps recommended by the World Health Organization (WHO) and a personalized handwashing method. ~\citet{zhang2019smart} introduces a smart ring incorporating an electrochemical fluid sensor. This sensor can detect the presence of various hand hygiene agents, such as tap water, soap, and hand sanitizer, and determine whether the handwashing duration meets the required standards for compliance.
For object-interacting activities, ~\citet{yamamoto2022ring} investigates the detection and classification of common hand-related activities in daily life. By simultaneously leveraging a proximity sensor on the ring and an IMU on the wristband to monitor wrist movements and finger flexion, they could distinguish six hand activities: eating, wiping, typing, writing, folding, and pegging. ~\citet{zhou2015healthcare}, on the other hand, uses the IMU on a ring to identify three fine motor activities: eating, brushing, and typing. Furthermore, several studies have pursued higher granularity in activity recognition, often requiring collaborative sensing with devices beyond the ring. ~\citet{moschetti2016recognition} leverages the IMU sensor on a ring to recognize nine gestures closely related to daily activities, such as eating with a hand, answering the telephone, etc. Even though this paper does not conduct experiments for daily activity recognition, their evaluation with 89\% classification accuracy on nine daily gestures shows a promising result.~\citet{liu2017novel} employs synchronized IMUs on both the wristband and the ring to recognize a variety of activities during eating and daily living, including cutting food with a fork and knife, opening a screw-top jar, taking the cap off a bottle and drinking, flipping pages of a magazine, buttoning a shirt, typing on a keyboard, and folding a towel. ~\citet{liang2019indexmo}  focuses their activity recognition on table-related activities, using RFID tags attached to both the fingers and utensils in addition to the ring IMU to detect activities such as touching the cup lid, holding the cup, lifting the plate, pouring water from a teapot, using a fork, using a knife, spooning water from the cup, and stirring water with a spoon.
~\citet{huang2019magtrack}  concentrates on driving behavior detection. Instead of IMUs, they use magnets in the ring and the wristwatch to sense hand movements. Magnets were also attached to the driver's glasses to detect fatigue-related behaviors. The activities detected in this study included left-hand position tracking, off-wheel detection, manual distraction detection, bimanual steering control monitoring, visual distraction detection, safe turning/lane changing, and fatigue motion detection.

\subsection{Fitness Tracking} \label{sec:outbody_fitness}
In addition to monitoring human activities at a granular level, smart rings also play a crucial role in fitness tracking to enhance user health management. Current research and applications of ring-based fitness tracking primarily focus on two key areas: estimating physical activity levels during the day and monitoring sleep quality during the night. 

~\citet{zhou2015healthcare} develops a healthcare system using the WondeRing device for activity detection. The analyzed data is stored and shared with supporters through an Android app, aiming to assist and ensure the well-being of elderly individuals living alone. ~\citet{friedman2014manumeter} uses a magnetic ring and wristband to measure wrist and finger movements, quantifying the daily use of the wrist and fingers through three metrics: wrist flexion/extension, wrist radial/ulnar deviation, and finger flexion/extension, thereby assessing limb usage in stroke patients. This approach offers a novel way to assess limb usage in stroke patients, potentially superseding questionnaires such as the Motor Activity Log (MAL) for evaluating post-stroke limb function.

Moreover, the insights gained from monitoring daily activities can extend to predicting sleep quality at night. 
Oura Ring can track daily activities and monitor various health parameters associated with sleep. ~\citet{kazemi2023can} conducts a longitudinal study involving 34 pregnant women over six months, utilizing the Oura Ring to continuously monitor their vital signs, physical activity, and sleep patterns. The study revealed that sedentary time significantly determines sleep duration, while inactive time exhibits a strong negative correlation with sleep efficiency. Based on these findings, the researchers developed an Extreme Gradient Boosting method to predict sleep quality attributes, leveraging daily activity data collected from wearable devices.

\subsection{Environment Recognition} \label{sec:outbody_environment}
The fingers are the most active body parts in daily interactions with the environment. Rings worn on the fingers are not only practical tools for monitoring personal activities but also for sensing the surrounding environmental cues. Some studies utilized the rings to gauge contextual information about the external world that the hands encounter. Additional research has explored the relationship between humans and the environment. 

\subsubsection{Environment Information Detection}
Smart rings equipped with multimodal sensors enable the detection of environmental information by capturing subtle physical or chemical properties of objects or surroundings. This capability facilitates context-aware interactions in ubiquitous computing scenarios. Hedgehog~\cite{yin2023hedgehog} is an innovative pervasive sensing methodology using optical sensors on a smart ring to detect drink spiking by analyzing light reflectivity changes due to particles in the drink to enhance personal safety. 


\subsubsection{Interaction Object Detection}
Beyond environmental sensing, smart rings further support detecting finger-object interactions, ranging from basic contact recognition to advanced characterization of interaction semantics (e.g., object type, interaction intent). This functionality is critical for enabling nuanced human-computer collaboration.  ~\citet{shi2020ready} leverages the differential micro-vibration signatures exhibited by fingertips in free space versus when in contact with objects, employing an IMU affixed to the fingertip to discern whether contact with an object has occurred.  
TextureSight~\cite{wang2024texturesight} is a smart ring that recognizes hand-engaged objects by detecting their unique surface textures using laser speckle imaging, enabling awareness of routine activities. Magic Finger~\cite{yang2012magic} is a ring device equipped with a micro RGB camera. Upon detecting contact between the finger and an object, the camera captures an image of the contact area. Machine learning algorithms then process the image, extracting feature vectors and classifying them to identify the texture of the contacted object. Magic Finger can recognize a variety of typical textures in both natural and artificial environments, such as tables, clothing, skin, and printed patterns, thereby triggering different actions.
~\citet{liang2019indexmo} extends this concept by developing a finger-worn RFID motion tracking system that detects objects through RFID and tracks hand movements using a smart ring to infer both the objects being interacted with and the nature of the interaction.
The aforementioned detection of finger-object interactions holds substantial potential for practical applications. ~\citet{kim2024iris} presents IRIS, a wireless, camera-enabled smart ring that allows users to interact with smart home devices by simply pointing at the target device and performing a corresponding gesture.
Z-Ring~\cite{waghmare2023zring} leverages bio-impedance sensing to detect interactions with objects. When a user touches an object, the hand’s electrical impedance changes, and these changes are captured by the ring’s sensors. The system then analyzes the impedance variations to identify the specific object being interacted with, enabling accurate object detection without requiring modifications to the objects themselves.

\subsection{Authentication} \label{sec:outbody_authentication}
Authentication via portable devices is an effective means of protecting user privacy and ensuring data security. 
Due to their small size, convenience, and continuous wearability, rings are an ideal choice for ubiquitous authentication compared to other portable devices.
Ring-based authentication can be implemented in several ways. One approach leverages the inherent patterns of hand movements detected by the ring, using these unique motion signatures to verify the user's identity. Another method treats the ring as a digital key. Finally, ring-based authentication can also be used in conjunction with other devices, focusing on verifying the user's identity when unlocking a device or entering sensitive information. 

\subsubsection{Behavioral Biometric Authentication}
Behavioral biometric authentication involves detecting behavioral characteristics of the user's hand motion through the ring.
Since each individual has distinct finger movement patterns—particularly during activities such as signing—finger motions tend to exhibit a consistent, idiosyncratic signature that differentiates one person from another. 
SigningRing ~\cite{cao2024signingring} is a system that utilizes sensor-based detection of signing movements for secure and rapid user authentication. When a user wearing the ring draws a signing pattern in the air, the system captures motion features such as speed, pauses, and intricate strokes. These features are then processed through a Siamese Neural Network (SNN) to determine their similarity to pre-registered patterns. Authentication is granted only if the captured features align with the registered ones. Pingu~\cite{roshandel2014multi} is a ring equipped with an IMU and a proximity sensor. Users wearing Pingu draw 3D signatures in the air or on a surface, and the device captures the corresponding sensor data. This data serves as a personalized signature template for each user. During authentication, Pingu compares the sensor data of the current gesture with the stored template. If the difference falls below a predefined threshold, the authentication is successful.

\subsubsection{Key-Based Authentication}
Each ring of the same model is assigned a unique identification code in this approach. Just like a traditional key, the correctness of the ring's information determines access, regardless of who possesses it. User authentication is achieved by verifying these ID codes during interactions, thereby ensuring that only the legitimate ring can access the associated services or devices. 

~\citet{vu2012distinguishing} introduces a novel authentication method using a ring that emits electrical signals. The ring has an onboard flash memory that stores bit sequences or messages, which can be user identifiers or secret keys for authentication. Upon contact with the screen, the ring sends electrical pulses through the wearer's finger, acting as a voltage source. These pulses trigger a series of artificial touch events on the screen, with the timing of these events corresponding to the bit sequence stored in the ring.~\citet{taylor2019finger} develops a wearable UHF RFID tag in the form of a ring for secure authentication. The ring incorporates a small loop antenna (SLA) that operates in the European UHF RFID frequency range. When illuminated by a UHF RFID reader, the ring can communicate over distances of up to 5 meters, depending on the substrate height. This design balances compactness and efficiency, making it ideal for securing identification applications.
OpenKeychain~\cite{schurmann2017openkeychain} is an Android-based cryptographic framework that enhances user authentication through the use of NFC rings. These rings securely store cryptographic keys on their embedded chips and communicate with the device via NFC to perform operations such as key generation, digital signatures, and encryption, all while keeping sensitive data isolated from the smartphone. By integrating PIN-based authentication with wearable technology, OpenKeychain achieves a balance between robust security and ease of use. The IR Ring~\cite{roth2010ir} is a wearable device in the form of a ring that emits a unique infrared signal to authenticate and track user interactions on a multi-touch display. Each ring produces a unique sequence of pseudorandom bits, which is detected and localized by the display's camera system. The display matches the received signal to the user's unique identifier, ensuring that all touch interactions near the signal are correctly attributed to the user.
~\citet{bianchi2017disambiguating} introduces a technique for touch disambiguation with a smart ring named VibRing. Users can send an invisible identifier to a touchscreen device by varying a distinctive vibration pattern on the ring. The device interprets the vibration pattern to recognize the user and their input, facilitating secure two-factor authentication and tailored interactions.

\subsubsection{Cross-Device Authentication}
Cross-device authentication involves verifying a user's identity when performing actions like unlocking devices or accessing sensitive information. This is achieved by correlating the motion signals from the ring with those of the other device to ensure consistency, thereby confirming that the ring's legitimate user is performing the actions.

PickRing~\cite{wolf2015pickring} is a wearable device that enables seamless interaction with electronic devices by detecting the motion of picking up a ring. It uses a ring-mounted gyroscope to capture motion data, which is transmitted to nearby devices via Bluetooth. The devices compare the received motion data with their own using an Euclidean distance calculation for pitch and roll, and they activate automatically when the patterns match, indicating that the device is being picked up. AuthoRing~\cite{liang2017authoring} is a wearable authentication system that continuously verifies user presence on computers. The ring collects hand motion data and sends it to the computer. When detecting mouse activity, the system compares this data with mouse movements using a correlation algorithm. This process identifies unauthorized users by detecting mismatches between the hand and mouse motions. AuthoRing prevents unauthorized access without needing hardware changes to the computer, leveraging existing BLE technology.

\section{Discussion}

Regarding the research question raised in Section~\ref{sec:methodology}, we have systematically examined \textbf{RQ1-3} in the preceding sections, and the key findings are summarized as follows:

\begin{itemize}
    \item \textbf{RQ1:} We have shown the distribution of smart ring papers in terms of publication years (Figure~\ref{fig:method_trend}) and proposed applications (Figure~\ref{fig:taxo_main}). Furthermore, we explore the distribution of papers across various subsets of applications, including gesture types (Table~\ref{tab:gesture_distribution}), commercial rings used (Table~\ref{tab:in-body_commercial}), and methods for establishing ground truth (Table~\ref{tab:health_gt_description}).
    
    \item \textbf{RQ2:} Our review demonstrates that approximately 87\% of the articles in our set of literature require a sensing solution for their proposed applications. We have comprehensively summarized the different sensor configurations in Table~\ref{tab:taxo_sensor} and described the phenomena that these sensors can detect in Table~\ref{tab:taxo_phenomena}.
    \item \textbf{RQ3:} To better understand the applications enabled by the identified sensing methods, we developed a four-layer taxonomy structured as ``$Application \rightarrow Phenomena \rightarrow Fundamental\,Phenomena (FPs) \rightarrow Sensors$'' as described in Section~\ref{sec:taxonomy}. We categorized applications into four primary groups using a ``binary decision tree'' approach: interaction - input, interaction - output, passive sensing - in-body feature, and passive sensing - out-body feature, as detailed in Section~\ref{sec:taxo_application}. Each study's focused application is further detailed in four corresponding sections.
\end{itemize}

We will now discuss the challenges and future directions for \textbf{RQ4} in this section.


\textit{Bridging the Gap: From Smart Ring Research to Real-World Adoption.}
Wearable devices have seen significant commercial success in recent years, particularly smartwatches, which have become mainstream in the consumer electronics market~\cite{Pangarkar2025smartwatch}. While smart rings have gained many traction, their adoption remains limited compared to smartwatches, despite their promising potential. This section explores the gaps between current smart ring research and their commercial viability, focusing on our categories of applications, along with associated challenges and future directions.

\subsection{Interaction - Input}
The unique placement of smart rings enables precise gesture-based input by detecting subtle finger movements, expanding the design space for gesture recognition. This allows for quick, intuitive interactions, such as controlling smart home devices~\cite{liu2023understanding}. Gesture input is especially useful in scenarios where traditional methods fail, like cooking with wet hands~\cite{han2016exploring} or holding objects~\cite{wolf2016microgesture}. However, seamless integration with other devices is essential, requiring a robust smart device ecosystem to fully unlock the potential of smart ring gestures, which limits the potential of smart ring gestures in daily activities.

Another promising application of smart ring inputs is their ability to mimic traditional input devices, such as keyboards (Section~\ref{sec:input_text_keyboard}), mice (Section~\ref{sec:input_traj_2d}), and VR controllers (Section~\ref{sec:input_traj_3d}). The smart ring's key advantage lies in constant availability and portability. However, further improvements are needed to enhance their efficiency and accuracy to match conventional input devices. For example, the most advanced smart ring virtual keyboard~\cite{gu2020qwertyring} currently achieves 20.35 words per minute (WPM), still lower than a standard computer keyboard. Additionally, existing smart ring input techniques impose certain operational constraints. Virtual keyboards often require fixed wrist or finger positions, such as QwertyRing~\cite{gu2020qwertyring} and RingVKB~\cite{li2023ringvkb}. Similarly, trajectory-based input methods limit motion space to improve precision, as seen in SoundTrak~\cite{zhang2017soundtrak} and MouseRing~\cite{shen2024mousering}. To ensure comfort and an always-available experience, factors such as ring size, the number of sensors, and power consumption are limited, which can be key reasons why some research cannot be practically implemented. Future advancements in adaptive algorithms and sensing techniques are crucial to overcoming these challenges, enabling more flexible, natural, and efficient smart ring inputs.

\subsection{Interaction - Output}

The majority of smart ring interaction outputs focus on haptic feedback (26 out of 29), as discussed in Section~\ref{sec:output_haptic}. One key challenge in smart ring haptics is integrating actuators into a compact form factor, particularly for skin-dragging~\cite{je2017designing, je2017tactoring} and poke-based haptics~\cite{je2018pokering}. Future advancements in miniaturization and actuation technologies could enhance the effectiveness and comfort of haptic feedback in smart rings. 

Additionally, there is limited comparative research on haptic feedback across wearable devices, such as smart rings and smartwatches, particularly in the context of notifications. Similarly, no studies have explored how smart rings compare to haptic gloves in VR/AR scenarios where portability is less critical. Future research could address these gaps by systematically evaluating the unique advantages of haptic feedback in smart rings. 

Another promising yet underexplored direction is leveraging smart rings to augment human perception, as discussed in Section~\ref{sec:output_enhance}. By integrating high-precision sensors and providing real-time perceptual feedback, smart rings could assist users in executing fine motor tasks~\cite{obushi2019magnifinger} and improving task efficiency~\cite{wang2012portable}.

\subsection{Passive Sensing}
\label{sec:discuss_open_platform}
Smart rings, known for their lightweight design, are widely used for health tracking, offering features such as physiological signal sensing, sleep monitoring, and fitness tracking. However, many passive sensing applications remain underexplored, particularly in disease diagnosis, treatment, authentication, and environmental sensing. These areas could greatly benefit from a smart ring platform that provides access to raw sensor data.

Currently, most smart rings offer physiological measurements like heart rate and SpO2 to developers, but these rely on proprietary inertial algorithms on the embedded system. The lack of access to raw sensor data and algorithmic details introduces potential biases, limiting researchers' ability to validate and refine measurements. Consequently, significant effort is required to ensure data reliability. Future advancements should focus on enabling controlled access to raw sensor data while maintaining privacy and security, fostering more robust and customizable health monitoring applications.

The research and development platforms for wearable technology play a crucial role in the effective utilization and integration of these devices in various applications. However, currently, there is a notable lack of robust, open platforms for the development of smart ring applications~\cite{zhou2023one}. On one hand, interaction rings offer unique possibilities for user interaction by interpreting gestures and motions as input commands. To harness these capabilities, it is essential that the development platform provides robust API support for integration with popular development environments. For instance, integrating interaction rings with Unity would allow developers to create more immersive and intuitive VR experiences. On the other hand, passive sensing rings are designed to collect data unobtrusively from the wearer's environment or physiological state. To fully exploit this capability, researchers need comprehensive APIs that allow for the extraction, processing, and interpretation of raw sensor data. This access allows developers to implement custom algorithms and applications that can provide more personalized and detailed insights into the data collected.

Further, future research should explore scalable deployment strategies to leverage smart rings in large-scale longitudinal studies~\cite{keeler2024biometrics}, addressing challenges such as data consistency, user adherence, and integration with existing healthcare systems. 

\subsection{Self-Containment}
Self-containment describes the ability to operate and compute without external supply. As rings are limited in size, they are incapable of incorporating heavy computing resources. Most previous work opted for external computational resources to function, while some research proposed alternatives for self-computing platforms \cite{magno2019self, gummeson2014energy}. However, as these platforms can be self-contained and support real-world applications, they are limited in sensor type and require energy harvesting methods, which restrain the usability of these prototypes. Generally speaking, higher self-containment requires trade-offs on sensing ability and operational time. To support in-the-wild usage of rings, the most common approach was to leverage wireless connection for data transition and use external devices for computation (e.g. laptops or smartphones). In these cases, rings are acting like a carrier of multiple sensors. As technology advances, smart rings could become more self-sufficient, featuring enhanced computational capabilities and extended battery life.

\subsection{Power Consumption and Battery Endurance.}
Due to the size and weight constraints of rings, making it difficult to accommodate large batteries, most ring prototypes in previous research relied on an external power supply. However, it is crucial to balance the ring's power consumption with battery endurance to enable extended wireless use in real-world conditions. Since sensor integration and cross-device communication are the primary contributors to power consumption, previous research \cite{boukhayma2021ring, chen2014mobiring, xu2022washring} has focused on the careful design of sensors (e.g. accelerator, low power PPG), reading circuits, and the choice of energy-efficient communication protocols (e.g. BLE). Moreover, multiple approaches were proposed to further reduce the power consumption of ring itself, including using low-power consumption sensors as triggers for functions, and passive reading \cite{takahashi2020telemetring, takahashi2024picoring}. On the other hand, battery endurance is another critical factor for achieving longer operational times. In addition to integrating larger batteries, various energy-harvesting methods have been proposed, such as using phones as wireless chargers or harnessing solar power \cite{nguyen2021smart, magno2019self, gummeson2014energy}. Given that rings are closely linked to users' touch interactions, there is potential to leverage wireless connectivity techniques to implicitly harvest power from nearby devices.



\subsection{User variability and culture}
In the development and evaluation of smart ring technologies, user variability and cultural differences play crucial roles in almost all applications summarized in this survey paper. The effectiveness of gesture recognition can vary significantly across users due to individual preferences~\cite{gheran2018gestures}, physical differences~\cite{altakrouri2016insights}, and cultural backgrounds~\cite{kwon2018cultural}. A gesture that is intuitive in one culture may be unfamiliar or even offensive in another, leading to inconsistencies in user interaction with the device. For physiological sensing technologies, photoplethysmography (PPG) for monitoring heart rate and blood oxygen levels can also be affected by user variability. Skin tone, for example, can influence the accuracy of PPG sensors, as darker skin may absorb more light, potentially leading to less accurate readings~\cite{scardulla2023photoplethysmograhic}. Activity recognition is another area where cultural differences are pronounced. Smart rings that aim to recognize and interpret specific activities must consider the diversity in how these activities are performed across different cultures. For example, eating habits vary widely: people might eat with their hands~\cite{moschetti2016recognition}, chopsticks, forks and knives~\cite{liu2017novel} in different culture.



\section{Conclusion}
In conclusion, this survey has comprehensively reviewed 206 pieces of literature pertaining to smart rings, categorizing them into four distinct application categories. Each category has been meticulously detailed in the corresponding sections of this paper. Furthermore, we introduced a phenomena-based taxonomy, which systematically breaks down the elements of smart ring technology into a hierarchy of applications, phenomena, fundamental phenomena, and sensors. Additionally, we have discussed several challenges that persist in the field, as well as potential directions for future research.

\begin{acks}
\end{acks}

\bibliographystyle{ACM-Reference-Format}
\bibliography{sample-base}

\appendix

\end{document}